\documentclass[11pt,letterpaper]{article}

\usepackage[utf8]{inputenc}
\usepackage[T1]{fontenc}
\usepackage{enumitem}
\usepackage{hyperref}
\usepackage{natbib}
\usepackage{url}
\usepackage{algorithm}
\usepackage[noend]{algpseudocode}
\usepackage{amsmath,amssymb}
\usepackage{amsthm}
\usepackage{cleveref}
\usepackage{booktabs}
\usepackage{graphicx}
\usepackage{bm}

\usepackage{libertine}
\usepackage[libertine]{newtxmath}

\usepackage[margin=1in]{geometry}
\usepackage[font={small,it}]{caption}

\usepackage{thmtools}
\usepackage{thm-restate}

\usepackage{dsfont}
\usepackage{mleftright}

\usepackage{tikz}
\usetikzlibrary{decorations.pathreplacing,angles,quotes}
\usetikzlibrary{shapes.geometric}

\allowdisplaybreaks 
\usepackage[normalem]{ulem}

\newcommand{\rsymb}{\textcircled{r}}

\newtheorem{example}{Example}

\newtheorem{theorem}{Theorem}

\newtheorem{lemma}{Lemma}[section]

\newtheorem{claim}{Claim}[section]
\newtheorem*{claim*}{Claim} 
\newtheorem*{proof*}{Proof}
\newtheorem{definition}{Definition}[section]
\newtheorem*{problem*}{Problem}

\newcommand{\expect}{\mathbb{E}}

\newcommand{\E}{\expect}

\newcommand{\opt}{\textrm{OPT}}

\newcommand{\I}{\mathcal{I}}
\newcommand{\M}{\mathcal{M} }

\newcommand{\abs}[1]{\lvert #1  \rvert}
\newcommand{\pr}[1]{\Pr \left( #1 \right)}

\newcommand{\fourapprox}{ \left( 4, O ( \log\log m  ) \right)}
\newcommand{\eightapprox}{\left( 8, O ( \log\log m \cdot \log k) \right)}
\newcommand{\mineleapprox}{\fourapprox}
\newcommand{\minbasiscardapprox}{\eightapprox}
\newcommand{\minkmatapprox}{\eightapprox}
\newcommand{\minkknapapprox}{\left(8, O\bigl(\log\log m \cdot \log^2{k} \bigr)\right)}
\newcommand{\minkknapthresh}{\left(1, O \bigl(\log{k}\bigr)\right)}
\newcommand{\mink}{\textrm{\textsc{Min}-}k}

\newcommand{\minbasis}{\textrm{\textsc{MinBasis}}}
\newcommand{\knapsack}{\textrm{\textsc{Knapsack}}}
\newcommand{\matroid}{\textrm{\textsc{Matroid}}}
\newcommand{\cardinality}{\textrm{\textsc{Cardinality}}}
\newcommand{\card}{\cardinality}
\newcommand{\minelement}{\textsc{Min-Element}}

\newcommand{\ex}[1]{\E \left[ #1 \right]}

\newcommand{\adapgreedy}{\textrm{\upshape{\textsc{AdapMGreedy}}}}

\definecolor{darkblue}{rgb}{.15,0,.7}
\hypersetup{
	colorlinks=true,
	linkcolor=darkblue,
	citecolor=darkblue,
	urlcolor=darkblue
}
\let\originalleft\left
\let\originalright\right
\renewcommand{\left}{\mathopen{}\mathclose\bgroup\originalleft}
\renewcommand{\right}{\aftergroup\egroup\originalright}
\newcommand{\p}[1]{\mathbb{P}\mleft(#1\mright)}

\newcommand{\policy}{\pi}
\newcommand{\set}{S}
\newcommand{\budget}{B}
\newcommand{\universe}{\mathcal{U}}
\newcommand{\alg}{\textrm{\upshape{\textsc{MetaMin}}}}
\newcommand{\resup}{R}

\newcommand{\threshapprox}{\textsc{Threshold}}

\newcommand{\intdomain}{[m]_{\geq 0}}
\newcommand{\given}{\;\middle|\;}

\DeclareMathOperator*{\argmax}{arg\,max}
\DeclareMathOperator*{\argmin}{arg\,min}

\newcommand{\leb}[1]{{\leq \frac{B}{#1}}}

\newcommand{\bucket}{\textsc{bucket}}
\newcommand{\ceiling}[1]{\lceil #1 \rceil}
\newcommand{\minknapsack}{\textrm{$\minelement$-$\knapsack$}}

\newcommand{\F}{\mathcal{F}}
\newcommand{\poly}{\textrm{poly}}

\newcommand{\expectb}[1]{\E \left[ #1 \right]}

\newcommand{\floor}[1]{\left \lfloor #1 \right \rfloor}

\newcommand{\adm}[1]{\textbf{Adm}\left( #1 \right )}

\newcommand{\weightfig}[1]{X_{ #1 }}
\newcommand{\AG}{\textrm{AG}}

\newcommand{\contract}[2]{#1 / #2}
\newcommand{\brk}[1]{\left \{ #1 \right \}}

\newcommand{\spi}{S^{(\pi)}}

\newcommand{\succprob}[2]{\pr{\trank{#1} \geq #2 }}
\newcommand{\Lpi}{L^{(\pi)}}
\newcommand{\LG}{L^{(G)}}
\newcommand{\Hsup}[1]{\bm{H}^{(#1)}}
\newcommand{\Rpi}{\Hsup{\pi}}
\newcommand{\RG}{\Hsup{G}}
\newcommand{\el}{\mathtt{l}}
\newcommand{\lpi}{\el^{(\pi)}}
\newcommand{\ppi}{p^{(\pi)}}
\newcommand{\cost}[1]{\mathtt{cost}\left( #1 \right)}

\newcommand{\width}[1]{ \mathtt{width}\left( #1 \right)}
\newcommand{\widthsum}[1]{g\left( #1 \right)}

\newcommand{\Tpi}{T^{(\pi)}}
\newcommand{\Tstar}{T^*}
\newcommand{\TG}{T^{(G)}}

\newcommand{\pitilde}{\tilde{\pi}}
\newcommand{\spitilde}{S^{(\pitilde)}}
\newcommand{\ssup}[1]{S^{\left( #1 \right)}}

\newcommand{\indc}[1]{\mathbf{1}_{\brk{#1}}}

\newcommand{\Minner}{\M_\mathbf{I}}
\newcommand{\Mouter}{\M_\mathbf{O}}

\newcommand{\ub}{\mathrm{UB}}
\newcommand{\mleftend}{a}
\newcommand{\mrightend}{b}

\newcommand{\rank}{\mathtt{rank}}
\newcommand{\trank}[1]{\rank\left( #1 \right)}

\newcommand{\pool}{\textsc{pool}}
\newcommand{\density}{a_e}
\newcommand{\bin}{\textrm{\upshape{\textsc{BIN}}}}
\newcommand{\greedyalg}{\textrm{\upshape{\textsc{ExtGreedy}}}}
\newcommand{\binset}{C}

\newcommand{\reward}[1]{\mathtt{reward}\left({#1} \right)}
\newcommand{\target}{i}

\newcommand{\spibi}{\spi_{\leb{i}}}
\newcommand{\targetnew}{\ell}

\newcommand{\matgreedy}{\text{\upshape{\textsc{MGreedy}}}}

\newcommand{\Mtrank}[2]{#1 \text{-} \trank{#2}}
\newcommand{\spix}{\ssup{\pi_x}}

\newcommand{\Ra}{R_a}
\newcommand{\Rb}{R_b}
\newcommand{\Rc}{R_c}

\renewcommand{\thefootnote}{\fnsymbol{footnote}}

\author{
Weina Wang\footnotemark{}\ \ {\rsymb} 
Anupam Gupta\footnotemark[\value{footnote}]\ \ {\rsymb} 
Jalani Williams\footnotemark[\value{footnote}] 
\footnote{Corresponding author. Email: \texttt{<jalaniw@cs.cmu.edu>}}
}

\footnotetext[1]{Computer Science Department, Carnegie Mellon University, Pittsburgh, PA 15213. }

\title{Probing to Minimize}
\date{}

\begin{document}

\maketitle

\begin{abstract}

We develop approximation algorithms for set-selection problems with deterministic constraints, but random objective values, i.e., stochastic probing problems.
When the goal is to maximize the objective, approximation algorithms
for probing problems are well-studied. On the other hand, few
techniques are known for minimizing the objective, especially in the adaptive setting, where information about the random objective is revealed during the set-selection process and allowed to influence it.
For minimization problems in particular, incorporating adaptivity can
have a considerable effect on performance. In this work, we seek
approximation algorithms that compare well to the optimal adaptive
policy.

We develop new techniques for adaptive minimization, applying them to
a few problems of interest. The core technique we develop here is an
approximate reduction from an adaptive expectation minimization
problem to a set of adaptive probability minimization problems which
we call \emph{threshold problems}. By providing near-optimal solutions
to these threshold problems, we obtain bicriteria \emph{adaptive}
policies.

We apply this method to
obtain an adaptive approximation algorithm for
the \textsc{Min-Element} problem, where the goal is to adaptively pick
 random variables to minimize the expected minimum value seen among
them, subject to a knapsack constraint. This partially resolves an open problem
raised in \cite{HowToProbe}. We further consider three extensions on the \textsc{Min-Element} problem, where our objective is the sum of the smallest $k$ element-weights, or the weight of the min-weight basis of a given matroid, or where the constraint is not given by a knapsack but by a matroid constraint. For all three of the variations we explore, we develop adaptive approximation algorithms for their corresponding threshold problems, and prove their near-optimality via
coupling arguments.

\end{abstract}
\renewcommand{\thefootnote}{\rsymb}
\footnotetext[0]{Denotes random author order, as seen \hyperlink{https://www.econometricsociety.org/publications/econometrica/information-authors/instructions-preparing-articles-publication}{here}.}
\renewcommand{\thefootnote}{\arabic{footnote}}
\thispagestyle{empty}

\newpage
\setcounter{page}{1}

\section{Introduction}\label{sec:introduction}

Consider the following stochastic optimization problem: given a
collection $X_1, X_2, \ldots, X_n$ of non-negative random variables,
with each r.v.\ $X_e$ having an associated cost $c_e$ and a known probability
distribution, pick a subset $S \subseteq [n]$ (from a collection of feasible sets
$\F \subseteq 2^{[n]}$) that minimizes
\[ f_{\min}(S) \triangleq \ex{\min_{e \in S} X_e}; \]
more generally, given a ``well-behaved'' function $f$, pick a set $S
\in \F$ to minimize
\[ \ex{f(S)}. \] Such a \emph{constrained stochastic minimization}
problem may arise in many contexts. 
For example, the government might have $\$100$ billion to disburse for vaccine research, and seek to produce a viable vaccine as quickly as possible. Given that each company $e$ requests a funding amount $c_e$ and, upon receiving funding, produces a viable vaccine in time $X_e$, the government might seek to fund companies in a way that minimizes the time it takes for at least one company to produce a viable  vaccine. 
Alternatively, consider a home-repair setting, where among a large pool of possible repairmen, a homeowner might be willing to ask for at most $k$ price estimates for a home-repair, and
seek to minimize her final price paid; or a collection setting, where among a pool of $4000$ possible sellers, a collector might purchase multiple copies of the same collectible from various distributors, and seek to minimize the number of defects in the best copy. 
In all of these settings, there is a natural
tradeoff between quality assurance and constraint impact which makes
these problems hard: often, riskier prospects are ``cheaper'' but
could pay off handsomely, while more stable prospects are either more
expensive or have a smaller possible upside.


While there has been considerable work on stochastic problems in the
\emph{maximization} setting (see, e.g., \cite{GuhaM-stoc07, adapsubmod, algsnadapgaps}), few algorithms and techniques are known
for \emph{minimization} problems. Indeed, the minimization problems
appear much harder.
One exception is the work of \citet{HowToProbe} who consider the $\minelement$ problem, 
in which we choose a set $S$ to minimize $\ex{\min_{e \in S}{X_e}}$,
subject to $S$ satisfying a knapsack constraint
$\sum_{e\in S} c_e \leq B$.  \citeauthor{HowToProbe}\ consider the
\emph{non-adaptive} version of this $\minelement$  problem, where one commits to a set $S$ of
variables before seeing any of the outcomes $\{X_e\}_{e \in S}$. 
They give a non-adaptive \emph{bicriteria} approximation to the
optimal non-adaptive policy, i.e., a set $S$ whose cost exceeds the budget
$B$ by an $O(\log\log m)$ factor (where the r.v.s take on values in
the set $\{0,\ldots, m\}$), such that $\E[ \min_{e \in S} X_e]$ is at
most $(1+\varepsilon)$ times the optimum. 

Such non-adaptive solutions are particularly interesting when they
have a small \emph{adaptivity gap}, i.e., when non-adaptive solutions have
a performance close to the best adaptive solutions. An \emph{adaptive}
solution builds its set $S$
element-by-element: the outcome of an r.v.'s weight is revealed
immediately after it is added to the selection set $S$, and the
selection policy can then \emph{adapt} and make future selections
based on all the outcomes seen so far. Unfortunately, simple examples
show that the $\minelement$ problem can have a large adaptivity gap:
see \S\ref{sec:appendix} for a case with three random
variables such that the best non-adaptive solution is arbitrarily
worse than the best adaptive one. 
Given such a large gap, it becomes
interesting to \emph{find adaptive solutions which compare well to the
  optimal adaptive policy}, so that we do not pay the price of the
huge adaptivity gap. \citeauthor{HowToProbe}\ mention getting good
adaptive solutions as a (potentially challenging) open problem; our
work does exactly this.


\subsection{Our results}

We focus on providing efficiently computable 
bicriteria guarantees.
We call an algorithm $\pi$ an
$(\alpha, \beta)$-approximation if the selection set $\spi$ is the
union of at most $\beta$ feasible sets, and the objective value
$\ex{f\left(\spi\right)}$ is at most $\alpha$ times the optimal expected objective value. 
Henceforth, $\intdomain \triangleq \{0,1,2,\dots, m \}$.

Our first main result is for the  
$\minelement$ problem in the adaptive setting; to differentiate it from the extensions we
consider
next, we call it the
$\minknapsack$ problem. 

\begin{restatable}{theorem}{RESTATEthmminele}
\label{thm:minelement}
For the $\minknapsack$ problem where the random variables take values in $\intdomain$, there exists an adaptive policy that obtains a $\mineleapprox$-approximation.
\end{restatable}

Theorem~\ref{thm:minelement} provides a partial answer to an open question in \cite{HowToProbe}.
Interestingly, our $O(\log\log{m})$ resource augmentation factor
\emph{matches} that obtained by \citeauthor{HowToProbe}\ in their
result for the non-adaptive setting. However, these factors seem to
arise for different reasons: their non-adaptive factor comes from a
use of greedy submodular optimization, while ours 
comes from a new \emph{adaptive binary search} procedure. This procedure,
which reduces the original adaptive
\emph{expectation minimization} problem to an adaptive
\emph{probability minimization} problem,
is a basic building block in our approach. 

The $\minknapsack$ can be generalized in many ways: we can 
consider \emph{richer function classes} (and not just the minimum function),
and we can consider \emph{richer constraint sets} (and not just the
simplest knapsack setting). In the first extension we present, 
the objective function
$f(S)$ is the sum of the smallest $k$ outcomes in $S$. We call this 
the $\mink$-$\knapsack$ problem. 

\begin{restatable}{theorem}{RESTATEthmminkknap}\label{thm:mink-knap}
    For the $\mink$-$\knapsack$ problem where the random variables take values in $\intdomain$, there exists an adaptive policy that obtains an $\minkknapapprox$-approximation.
\end{restatable}
To prove Theorem~\ref{thm:mink-knap}, we 
use the same adaptive binary search idea from
Theorem~\ref{thm:minelement}, 
and relate the resultant probability minimization problem to a different probability maximization problem
that we call the \emph{$i$-heads problem}, which is an interesting problem in its own right.
We then obtain a better-than-optimal solution for this problem by extending
the simple greedy algorithm used for Theorem~\ref{thm:minelement}, albeit 
overspending by an $O(\log{k})$ factor. A detailed discussion appears
in the techniques section.

Next we consider 
the $\mink$-$\matroid$ problem, where the objective function $f(S)$ is
again the sum of the smallest $k$ outcomes in the selection set $S$,
but the constraint is now a matroid constraint. \footnote{A
matroid $\Mouter = (\universe, \I)$ 
is specified by a ground set $\universe$ and a family of independent sets $\I$. 
A selection set $\spi$ is feasible if and only if it is an independent
set, i.e., $\spi$ lies in the family of feasible sets $\F = \I$.}
Our solution for this setting has an even better approximation than
for the case of knapsack constraints.
\begin{restatable}{theorem}{RESTATEthmminkmat}\label{thm:mink-mat}
    For the $\mink$-$\matroid$ problem where the random variables take values in $\intdomain$, there exists an adaptive policy that obtains a $\minkmatapprox$-approximation.
\end{restatable}

Theorem~\ref{thm:mink-mat} takes the techniques developed for the
$\mink$-$\knapsack$ problem, and shows the core reason why those
techniques work: framed 
correctly, both constraints admit a nice sense of interchangeability. For the adaptive probability minimization problem we reduce to here, a \emph{non-adaptive} greedy algorithm actually becomes \emph{optimal}.


Finally, we investigate the
$\minbasis$-$\card$ problem, where the constraint is a cardinality
constraint (we can pick at most $B$ elements), but we
\emph{generalize} the objective function from being the sum of the
smallest $k$ random variables to the setting of
matroids. Specifically, we now have a matroid
$\Minner=(\universe,\I)$ of rank $k$, and we
consider 
the minimum-weight basis in this matroid.
One should note that taking a 
uniform matroid as the constraint here gives us back the $\mink$-$\knapsack$ problem, albeit in the case where all items have unit cost.


\begin{restatable}{theorem}{RESTATEthmminbasiscard}\label{thm:minbasis-card}
    For the $\minbasis$-$\cardinality$ problem where the random variables takes value in $\intdomain$, there exists an adaptive policy that obtains a $\minbasiscardapprox$-approximation.
\end{restatable}

While the result of Theorem~\ref{thm:mink-mat} shows the importance of interchangeability
in the constraint, Theorem~\ref{thm:minbasis-card} shows the importance of
interchangeability in the objective. When the matroid constraint moves
into the objective, the previously optimal non-adaptive matroid greedy
algorithm now needs to be made \emph{adaptive} in order to maintain
optimality. We note a peculiarity here: while solving the non-adaptive
probability minimization problem is non-trivial, the \emph{adaptive}
version of this problem has a simpler optimal solution.

\subsection{Techniques}\label{sec:techniques}

We now give some more details about the main techniques behind our 
results. 

\subsubsection{Reduction to threshold problems}
\label{sec:reduct-thresh-probl}

The primary technique we use is a reduction to threshold problems. The
idea is a clean one, provided we observe that 
\begin{equation}
\ex{f\left(\spi\right)}\le\pr{f\left(\spi\right)>0}+\sum_{j=0}^{\lfloor
  \log m\rfloor}\pr{f\left(\spi\right)>2^j}2^j \le
2\ex{f\left(\spi\right)}. \label{eq:1}
\end{equation}
This suggests that 
a good policy for minimizing $\ex{f\left(\spi\right)}$ should somehow simultaneously minimize
all of these $\pr{f\left(\spi\right)>t}$ terms
for values of $t$ that are powers of $2$.  
Now if we have an algorithm, referred to as $\threshapprox$, which can $(\alpha, \beta)$-approximate
these 
threshold problems for every power-of-$2$ threshold, we obtain a
$\left(2\alpha, O(\log{m})\beta\right)$-approximation to the optimal
adaptive policy. But this is incredibly wasteful, and does not use the
adaptivity.  
To avoid this loss, we use the property that $f(\cdot)$ is non-increasing; if we
obtain $f\left(\spi\right) \leq t$, we no longer need to worry about solving threshold
problems with thresholds larger than $t$.  To make use of this
observation, we instead perform an \emph{adaptive binary search} to
determine the next threshold to call $\threshapprox$ on.



\medskip\noindent\textbf{Adaptive binary search.} 
Starting with the median threshold $t$, $\threshapprox$ probes
a set $S_t$ that aims to minimize the probability of $f(S_t) >
t$. If $\threshapprox$ obtains $f(S_t) \le
t$, we say it succeeds and we recurse on the thresholds smaller than $t$; if
$\threshapprox$ fails to obtain $f(S_t) \leq t$, we recurse on the set
of thresholds larger than $t$. While this seems simple, one should
note the asymmetry between success and failure here: success at a
threshold $t$ guarantees that our objective value is at most  $t$ from then on, but failure
at $t$ \emph{does not guarantee that the objective value is greater than $t$ from then on}.
Nevertheless, we construct an upper bound $\ub$ on our algorithm's objective value, which allows us to
analyze it as if failure at $t$ \emph{does} imply that the objective is at least $t$ from then on. See \S\ref{sec:reduction} for details.  This
adaptive binary search now reduces the number of calls to
$O(\log \log m)$. 

\subsubsection{\texorpdfstring{The $\minknapsack$ problem}{The MIN-ELEMENT-KNAPSACK problem}}
\label{sec:minkn-pr}

Armed with this reduction 
to a threshold problems, 
the proof of Theorem~\ref{thm:minelement} follows from an 
approximation to the $\minknapsack$ threshold problem
\begin{equation*}
    \min_{\pi \in \adm{\F}}{\pr{\min_{e \in \spi}{X_e} > t}}.
\end{equation*}
Since any policy $\pi$ succeeds at obtaining $\min_{e\in S}{X_e} \leq
t$ as soon as it finds a single element for which $X_e \leq t$, 
adaptivity adds nothing to solving this problem. Hence we focus 
on the non-adaptive problem
\begin{equation*}
    \min_{S \in \F}{\pr{\min_{e\in S}{X_e} > t}}.
\end{equation*}
Taking the logarithm of the objective, 
we reduce to a knapsack instance where each element has \emph{reward} equal to  $-\log{\pr{X_e > t}}$.
Now a greedy procedure gives a $(1,2)$-approximation to the threshold
problem. Combining all these together gives the 
$\mineleapprox$-approximation from Theorem~\ref{thm:minelement}.

\subsubsection{\texorpdfstring{The $\mink$-$\knapsack$ problem}{The MinK-Knapsack problem}}\label{sec:intro-mink}

We turn next to a strict extension of the $\minknapsack$ problem: the
$\mink$-$\knapsack$ problem. The first step, like in the
$\minknapsack$ problem, is to apply a threshold-based reduction.  An
idea similar to that from \S\ref{sec:reduct-thresh-probl} allows us to
reduce a slightly richer set of objective functions
to threshold minimization problems. 
Indeed, if $y_i(S)$ is the weight of the $i$-th smallest weight
element in $S$, then our objective $f(S)$ is $\sum_{i=1}^k{y_i(S)}$,
the sum of the $k$ smallest weights in $S$. Writing the expectation as
a sum of thresholds like in~(\ref{eq:1}), we get that it suffices to 
find approximations to the threshold problem
\begin{equation*}
  \min_{\pi \in \adm{\F}}{\pr{y_i\left(\spi\right) > t}}
\end{equation*}
for different values of $t$.  This problem 
can be thought of in the following way.
\begin{problem*}[\textbf{The $i$-heads problem}]
  We are given a set of coins $\universe$, each having cost $c_e$ and
  bias $p_e$. We can flip a coin $e$ at most once, at a cost of
  $c_e$. Given a budget $B$, the $i$-heads problem seeks to adaptively
  flip coins to maximize $\pr{\text{see at least $i$ heads}}$.
\end{problem*}


We extend the greedy knapsack strategy from \S\ref{sec:minkn-pr}
used to solve the case $i =1$ to the case of a general $i$: the idea
is again to pick items in increasing order of cost-per-reward, except
that the budget is extended from $B$ to $B + i\delta$, where $\delta$
is the maximum cost of any element. In order to control this $\delta$
term, we bucket the coins and choose the best coins at each of the top
$O(\log i)$ levels until the cost of each coin becomes smaller than
$B/i$.



The proof that this greedy policy
$G$ 
outperforms any admissible policy $\pi$ requires some new ideas. The
heart of the argument is to first construct a surrogate policy
$\pitilde$ which strictly outperforms any policy $\pi$, and then to
show  
an ordering of the coins of $G$ such that
\begin{itemize}
    \item The policies $G$ and $\pitilde$ probe the same set after obtaining their first heads.
    \item For any amount of remaining budget $B'$, the probability that $G$ obtains its first heads with $B'$ budget remaining is greater than the probability that $\pi$ obtains its first heads with $B'$ budget remaining.
\end{itemize}
In the end, our argument shows that $G$ stochastically dominates every
admissible policy $\pi$, which completes the proof.




\subsubsection{\texorpdfstring{The $\mink$-$\matroid$ problem}{The MinK-Matroid problem}}

We now discuss our first matroid-based variation on the
$\mink$-$\knapsack$ problem, which we call the $\mink$-$\matroid$ problem. 
The same reduction used in the $\mink$-$\knapsack$ problem implies that
proving Theorem~\ref{thm:mink-mat} reduces to providing a $(1, O(1))$-approximation to 
threshold problem for matroids. 
We show that the matroid greedy algorithm $\matgreedy$, described in Algorithm~\ref{alg:mink-mat}, is an optimal policy for this problem. In fact, the matroid greedy algorithm stochastically dominates any other policy. We argue this via induction on the matroid rank. The main observation one must make here is that, after probing a single element, we find ourselves in a problem state nearly exactly like the starting state, but with a possibly reduced number of heads, and a resulting constraint family which is a matroid with a strictly smaller rank.

\subsubsection{\texorpdfstring{The $\minbasis$-$\cardinality$ problem}{The MinBasis-Cardinality problem}}
We now discuss the treatment of our final problem, the
$\minbasis$-$\cardinality$ problem. The proof of
Theorem~\ref{thm:minbasis-card} is again through the 
reduction to threshold problems for an objective $f(S)$ that is in the form of $\sum_{i=1}^k{g_i(S)}$.
However, unlike the case of a $\mink$ objective, now it is not obvious how we can
write the $\minbasis$ objective $f(S)$ as $\sum_{i=1}^k{g_i(S)}$ for a monotonic
function sequence $(g_i(\cdot))$ where each $g_i(\cdot)$ is a
non-decreasing function taking values in $\intdomain$.  In our proof,
we first show that we can specify $g_i(S)$ to be the $i$-th smallest
element in the minimum-weight basis contained in $S$ generated by the standard
matroid greedy algorithm.  Then we show inductively that
$\adapgreedy$, a simple adaptive matroid greedy algorithm described in
Algorithm~\ref{alg:minbasis-card}, is optimal for the resulting
threshold problem.  The key idea here is again
to use our inductive assumption to reason about the future behavior of
an optimal policy.




\subsection{Related Work}

The work closest to ours in content and aim is \cite{HowToProbe}. 
That paper allows a more general model, where element weights can take on non-negative discrete values $\mu_1, \mu_2, \dots, \mu_L$, but restricts it focus to the $\minelement$ objective $f(S) \triangleq \min_{e\in
  S}{X_e}$.
\citeauthor{HowToProbe}\ give a bicriteria
approximation scheme, showing for every $\varepsilon > 0$ how to obtain a set $T$ such that
\begin{equation*}
    \cost{T} \leq O\left(\log(L) + \log{\frac{1}{\log{(1+\varepsilon)}}}\right)
\end{equation*} 
and meanwhile $\ex{f(T)} \leq (1+\varepsilon) \ex{f(S^*)}$ for $S^* \triangleq \argmin_{S \in
  \F}{\ex{f(S)}}$. Moreover, they
show that finding a $\poly(L)$ 
approximation to the non-adaptive
problem is NP-hard, by establishing an approximation-preserving
reduction to a special class of covering integer programs. 
Our results differ in several ways. 
First, \citeauthor{HowToProbe}\ consider non-adaptive approximations
to the optimal non-adaptive policy, whereas
we provide adaptive approximations to the optimal adaptive policy,
partially closing an open question posed by them. 
Second, they obtain a $(1+\varepsilon)$-approximation instead of a
constant-factor approximation. 
Third, their
techniques are quite different from ours: they use the submodularity
of (a transformation of) the expected objective $\E[f(S)]$, while we adaptively explore elements through an adaptive binary search.
Moreover, we consider several generalizations of the $\minelement$ problem with more general objectives and constraints.

Stochastic \emph{maximization} problems have been studied much more
broadly: the results for these problems very often show small
adaptivity gaps and focus on particular classes of objective functions \citep{AdamSW13, XOS, StochMVProbing, adapsubmod}. In contrast, we show that one of our minimization problems
has a large adaptivity gap, and our main technical result, the reduction to threshold problems, applies to a general class of objective functions. Moreover, the approaches used for these
problems and for ours are quite different.

With respect to solving threshold problems, to the authors’ knowledge, generally little is known. However, the threshold problem associated with the $\mink$-$\knapsack$ problem (see \S~\ref{sec:mink-knap}) has been heuristically treated before. In particular, it is a Bernoulli version of a more general adaptive knapsack problem, where rewards are independently random but the cost of each item is fixed, and the goal is to achieve a cumulative reward above some threshold with optimal probability. Previous work (\cite{targetachieve}, \cite{adapknap}) on this problem has focused on the development of heuristics and on optimizing the dynamic programming solution, in the setting where rewards are normally distributed. This work is instead concerned with the development of bicriteria approximation algorithms in the setting where rewards are Bernoulli distributed, though it may be possible to extend our methods to the more general case.


\section{General Framework with Threshold Problems}\label{sec:reduction}

In this section, we present the full details of the reduction from the stochastic minimization problem to threshold problems, first alluded to in Section~\ref{sec:techniques}.
Formally, the $f$-$\F$  threshold problem with threshold $t$  is an adaptive probability minimization problem, wherein the policy $\pi$ tries to avoid $f(\spi) > t$ with as high a probability as possible, while abiding by the constraint that $\spi \in \F$. Lemma~\ref{lem:reduction} says that, given an $(\alpha, \beta)$-approximation for the $f$-$\F$ threshold problem, one can construct a $\left(4\alpha, O(\log\log{m})\beta  \right)$-approximation for the adaptive expectation minimization problem $\min_{\pi \in \adm{\F}}{\ex{f(\spi)}}$.
Our results are summarized in Lemma~\ref{lem:reduction} and Corollary~\ref{coro:sum-of-k}, which we state now.



\begin{restatable}[\textbf{Reduction to threshold problems}]{lemma}{RESTATElemreduction}\label{lem:reduction}
Let $f(S)$ be a non-increasing objective function taking values in $\intdomain$ for a positive integer $m$. Let $\F$ be the constraint family and $\adm{\F}$ be the set of feasible policies under $\F$.  Suppose that for any $t \in \intdomain$, the $f$-$\F$ threshold problem
\begin{equation*}
    \min_{\pi \in \adm{\F}}{\pr{f\left(\spi\right) > t }}
\end{equation*}
admits an $(\alpha,\beta)$-approximation.  Then the stochastic minimization problem 
\begin{equation*}
\min_{\pi \in \adm{\F}}{\ex{f\left(\spi\right)}}
\end{equation*}
admits a $(4\alpha, O(\log\log m)\cdot \beta)$-approximation.
\end{restatable}

\begin{restatable}[\textbf{Sum-of-$k$ reduction}]{coro}{RESTATEcorosumofk}
\label{coro:sum-of-k}
    Let $g_i(S)$, $i \in [k]$, be a non-increasing function taking values in $\intdomain$, and assume that the function sequence $(g_i)$ is monotonic in $i$.  Let $f(S) = \sum_{i=1}^k{g_i(S)}$. Suppose that for any $t \in \intdomain$, the $g_i$-$\F$ threshold problem 
    \begin{equation*}
    \min_{\pi \in \adm{\F}}{\pr{g_i\left(\spi\right) > t }}
    \end{equation*}
    admits an $(\alpha, \beta)$-approximation. Then the stochastic minimization problem 
    \begin{equation*}
    \min_{\pi \in \adm{\F}}{\ex{f\left(\spi\right)}}
    \end{equation*}
    admits an $(8\alpha,  O(\log\log m\cdot \log k) \cdot \beta)$ approximation.
\end{restatable}

As described in Section~\ref{sec:techniques}, to build the intuition for the reduction, we start with the following decomposition for any policy~$\pi$:
\begin{equation}\label{eq:decomp}
\ex{f\left(\spi\right)}\le\pr{f\left(\spi\right)>0}+\sum_{j=0}^{\lfloor \log m\rfloor}\pr{f\left(\spi\right)>2^j}2^j \le 2\ex{f\left(\spi\right)}.
\end{equation}
This decomposition follows from the relation $\ex{f\left(\set^{(\policy)}\right)}=\sum_{t=0}^{m}\pr{f\left(\set^{(\policy)}\right)>t}$ and the following inequalities for any non-increasing non-negative sequence $(a_t)_{t=0}^{m}$ by taking $a_t=\pr{f\left(\set^{(\policy)}\right)>t}$:
\begin{equation}\label{eq:condensation}
    \sum_{t=0}^{m}{a_t} \leq a_0 + \sum_{j=0}^{\floor{\log m}}{a_{2^j} 2^j} \leq 2 \sum_{t=0}^{m}{a_t},
\end{equation}
which is a slightly modified statement of the inequalities used in the Cauchy condensation test of \cite{rudin1976principles}.

With the decomposition in \eqref{eq:decomp}, consider the following policy $\widehat{\pi}$.
Let $\widehat{S}_{t}$ be the $(\alpha,\beta)$-approximate solution of the threshold problem $\min_{\pi \in \adm{\F}}{\pr{f\left(\spi\right) > t }}$, and let $S^{(\widehat{\pi})}=\widehat{S}_{0}\cup \widehat{S}_{2^0}\cup \widehat{S}_{2^1}\cup\cdots\cup \widehat{S}_{2^{\lfloor \log m\rfloor}}$.
Then
\begin{align*}
\ex{f\left(S^{(\widehat{\pi})}\right)}
&\le\pr{f\left(S^{(\widehat{\pi})}\right)>0}+\sum_{j=0}^{\lfloor \log m\rfloor}\pr{f\left(S^{(\widehat{\pi})}\right)>2^j}2^j\\
&\le\pr{f\left(\widehat{S}_{0}\right)>0}+\sum_{j=0}^{\lfloor \log m\rfloor}\pr{f\left(\widehat{S}_{2^j}\right)>2^j}2^j\\
&\le \alpha\pr{\opt>0}+\sum_{j=0}^{\lfloor \log m\rfloor}\alpha\pr{\opt>2^j}2^j\\
&\le 2\alpha\ex{\opt},
\end{align*}
where $\opt$ is the value of an optimal policy for $\min_{\pi \in \adm{\F}}{\ex{f\left(\spi\right)}}$, and we have used the fact that $f(\cdot)$ is non-increasing.
Unfortunately, the policy $\widehat{\pi}$ uses a resource augmentation factor of the order of $\log m$.
In the proof of Lemma~\ref{lem:reduction}, we further reduce the resource augmentation factor using the policy $\alg$, described in Algorithm~\ref{alg:extreme}, which performs an adaptive binary search on the threshold values $R=\{0,2^0,2^1,\dots,2^{\lfloor \log m \rfloor}\}$.

\begin{algorithm}
\caption{$\alg (\F)$}
\label{alg:extreme}
\begin{algorithmic}[1]
\State Initialization: $\resup\gets\{0\}\cup\{2^j\colon j=0,1,\dots,\lfloor \log m\rfloor\}$; $\widehat{\set}\gets \emptyset$
\State $\widehat{\set}_0\gets \threshapprox(\F, 0)$ \Comment{\textbf{Boundary case}}
\State $\widehat{\set}\gets\widehat{\set}\cup\widehat{\set}_0$
\If{$f\mleft(\widehat{\set}_0\mright)=0$} \Comment{Succeeds the threshold test at $0$}
	\State\Return $\widehat{\set}$
\Else \Comment{Fails the threshold test at $0$}
	\State $\resup\gets\resup\setminus\{0\}$
\EndIf
\While{$\resup\neq \emptyset$} \Comment{\textbf{Adaptive threshold testing}}
	\State $t\gets$ median of $\resup$
	\State $\widehat{\set}_t\gets \threshapprox(\F,t)$
	\State $\widehat{\set}\gets\widehat{\set}\cup\widehat{\set}_t$
	\If{$f\mleft(\widehat{\set}_t\mright)\le t$} \Comment{Succeeds the threshold test at $t$}
		\State $\resup\gets\resup\setminus\{\tau\in\resup\colon \tau\ge t\}$
	\Else \Comment{Fails the threshold test at $t$}
		\State $\resup\gets\resup\setminus\{\tau\in\resup\colon \tau\le t\}$
	\EndIf
\EndWhile

\State \Return $\widehat{\set}$
\end{algorithmic}
\end{algorithm}

\subsection{Proof of Lemma~\ref{lem:reduction} (Reduction to threshold problems)}
\begin{proof}
We prove Lemma~\ref{lem:reduction} by showing that the policy $\alg$ in Algorithm~\ref{alg:extreme} is a $(4\alpha, O(\log\log m)\cdot \beta)$-approximation. It is easy to see that the final set $\ssup{\alg}$ is the union of at most $\left(O(\log\log{m})\cdot \beta\right)$ $\F$-feasible sets, since $\alg$ performs a binary search on the threshold values $R=\{0,2^0,2^1,\dots,2^{\lfloor \log m \rfloor}\}$.
Therefore, it suffices to show that $\ex{\alg} \leq 4 \alpha \ex{\opt}$.

We first construct an upper bound $\ub$ on the objective value $\alg$ whose value directly reflects the failing or succeeding of threshold tests.
Specifically, let $\tau$ be the last threshold at which $\alg$ fails, and we define
\begin{equation}
    \ub = \max \{\alg, \tau + 1\}.
\end{equation}
Clearly, $\alg\le \ub$ by definition.
Note that $\tau$ is the largest threshold at which $\alg$ fails due to the binary search procedure.
So we can guarantee that $\ub$ cannot achieve a value smaller than or equal to a threshold value $t$ if $\alg$ fails the test at $t$.

We now show how the constructed upper bound $\ub$ reflects the test results.
For notational convenience, we partition the interval $[0,m]$ into the following intervals:
$$[0,m]=\{0\}\cup (0,1]\cup(1,2]\cup(2,2^2]\cup\dots\cup(2^{\lfloor \log m\rfloor-1},2^{\lfloor \log m\rfloor}]\cup(2^{\lfloor \log m\rfloor},m],$$
which we denote as
\begin{align*}
\I_{-1}&=(0,1]\triangleq(\mleftend_{-1},\mrightend_{-1}],\\
\I_{j}&=(2^j,2^{j+1}]\triangleq(\mleftend_j,\mrightend_j],\;j=0,1,\dots,\lfloor\log m\rfloor-1,\\
\I_{\lfloor\log m\rfloor}&=(2^{\lfloor\log m\rfloor},m]\triangleq(\mleftend_{\lfloor\log m\rfloor},\mrightend_{\lfloor\log m\rfloor}].
\end{align*}
We use the phrase that ``$\alg$ succeeds/fails the test at a threshold'' to mean that ``$\alg$ indeed \emph{performs} the test and \emph{succeeds/fails}''.

The key to our proof is the claim below.  Note that such a relation does not hold for $\pr{\alg \in (\mleftend_j,\mrightend_j]}$.
It is possible that $\alg$ fails at some $\mleftend_j$ but ultimately attains a value smaller than $\mleftend_j$ since it happens to observe a small $f\bigl(\widehat{S}_t\bigr)$ when it succeeds at another threshold~$t$.
\begin{claim*}
For any $j\in\{-1,0,1,\dots,{\lfloor \log m\rfloor}\}$,
\begin{equation*}
\pr{\ub \in (\mleftend_j,\mrightend_j]}=\pr{\alg \text{ fails at $\mleftend_j$ and succeeds at $\mrightend_j$}}.
\end{equation*}
\end{claim*}
\begin{proof}[Proof of Claim.]
It suffices to focus on the case where $\alg$ fails the test at threshold $0$.  
We first argue that if $\alg$ fails at $\mleftend_j$ and succeeds at $\mrightend_j$, then $\ub\in(\mleftend_j,\mrightend_j]$.
By the construction of $\ub$, we have seen that if $\alg$ fails the test at $\mleftend_j$, then $\ub\ge \mleftend_j+1>\mleftend_j$.
If $\alg$ succeeds the test at $\mrightend_j$, then $\alg\le \mrightend_j$, and $\alg$ does not perform tests at threshold values above $\mrightend_j$ and thus does not have a chance to fail at these threshold values.
Therefore, $\ub\le \mrightend_j$.

We now show that if $\ub \in (\mleftend_j,\mrightend_j]$, then $\alg$ fails at $\mleftend_j$ and succeeds at $\mrightend_j$.
After $\alg$ terminates, let $t_F$ be the largest threshold that $\alg$ has failed at, and let $t_S$ be the smallest threshold that $\alg$ has succeeded at.
We argue that it must hold that $t_F=\mleftend_j$ and $t_S=\mrightend_j$, which is sufficient for the claim.
By the termination condition, $(t_F,t_S]$ must be an interval among the class of intervals $\{(\mleftend_u,\mrightend_u]\colon u=-1,0,1,\dots,\lfloor\log m\rfloor\}$.
By the argument in the last paragraph, the fact that $\alg$ fails at $t_F$ and succeeds at $t_S$ implies that $\ub\in(t_F, t_S]$.
Since the intervals $(\mleftend_u,\mrightend_u]$'s are disjoint, it must be the case that $(t_F, t_S]$ is the same interval as $(\mleftend_j,\mrightend_j]$, which completes the proof.
\end{proof}

With this claim, we complete the proof of Lemma~\ref{lem:reduction} through the following inequalities.
First, it is easy to see that the claim implies that
\begin{align*}
\pr{\ub\in (\mleftend_j,\mrightend_j]}\le \pr{\alg \text{ fails at }\mleftend_j}\le\pr{f\mleft(\widehat{\set}_{\mleftend_j}\mright) > \mleftend_j},
\end{align*}
where $\widehat{\set}_{\mleftend_j}$ is the set chosen by $\threshapprox$ for the threshold problem at threshold~$\mleftend_j$.
Then
\begin{align*}
\ex{\ub}&\le\sum_{j=-1}^{\lfloor \log m\rfloor}\pr{\ub\in (\mleftend_j,\mrightend_j]}\cdot\mrightend_j\\
&\le\sum_{j=-1}^{\lfloor \log m\rfloor}\pr{f\mleft(\widehat{\set}_{\mleftend_j}\mright) > \mleftend_j}\cdot 2^{j+1}\\
&\le \sum_{j=-1}^{\lfloor \log m\rfloor}\alpha\pr{\opt > \mleftend_j}\cdot 2^{j+1}\\
&=\alpha\pr{\opt > 0}+2\alpha\sum_{j=0}^{\lfloor \log m\rfloor}\pr{\opt>2^j}\cdot 2^{j}\\
&\le 4\alpha\ex{\opt},
\end{align*}
where the last line follows from the decomposition in \eqref{eq:decomp}.  Recalling that $\ex{\alg}\le\ex{\ub}$, we have completed the proof that $\alg$ is a $(4\alpha, O(\log\log m)\cdot \beta)$-approximation.
\end{proof}

\subsection{\texorpdfstring{Proof of Corollary~\ref{coro:sum-of-k} (Sum-of-$k$ reduction)}{Proof of Corollary~\ref{coro:sum-of-k} (Sum-of-k reduction)}}
\begin{proof}
Without loss of generality, we assume that the function sequence $(g_i)$ is non-increasing in $i$ since otherwise we can reverse their indices.
Applying the powers-of-$2$ condensation in \eqref{eq:condensation} for the non-increasing sequence $g_1(S),g_2(S),\dots,g_k(S)$, we have that for any policy $\pi$,
\begin{equation*}
    \expectb{\sum_{i=1}^{k}{g_i\left(\spi\right)}}  \leq \ex{ \sum_{j=0}^{\floor{\log(k)}}{g_{2^j}\left(\spi\right)2^j}}\le 2\expectb{\sum_{i=1}^{k}{g_i\left(\spi\right)}}.
\end{equation*}

We now consider the following policy $\widetilde{\pi}$.  For each $j=0,1,\dots,\floor{\log(k)}$, let $S_j$ be the output of the policy $\alg$ in Algorithm~\ref{alg:extreme} for
\begin{equation*}
\min_{\pi \in \adm{\F}}{\ex{g_{2^j}\left(\spi\right)}}.
\end{equation*}
Then let $S^{(\widetilde{\pi})}=S_0\cup S_1\cup \dots \cup S_{2^{\floor{\log k}}}$.
Recall that $\alg$ is a $(4\alpha, O(\log\log m)\cdot \beta)$-approximation.
Clearly, the policy $\widetilde{\pi}$ has an augmentation factor of $O(\log\log m\cdot \log k)\cdot \beta$.
Further,
\begin{align*}
\expectb{\sum_{i=1}^{k}{g_i\left(S^{(\widetilde{\pi})}\right)}}
&\leq \sum_{j=0}^{\floor{\log(k)}}\ex{{g_{2^j}\left(S^{(\widetilde{\pi})}\right)}}\cdot 2^j\\
&\leq \sum_{j=0}^{\floor{\log(k)}}\ex{{g_{2^j}\left(S_j\right)}}\cdot 2^j\\
&\leq \sum_{j=0}^{\floor{\log(k)}}{4 \alpha \ex{g_{2^j}\left(S^{(\opt)}\right)}}\cdot 2^j\\
&\leq 8 \alpha \ex{\opt}.
\end{align*}
This completes the proof that the policy $\widetilde{\pi}$ is a $\left(8\alpha, O(\log\log m\cdot \log k)\cdot \beta\right)$-approximation.
\end{proof}

\section{\texorpdfstring{The $\minknapsack$ Problem}{The Min-Element-Knapsack Problem}}\label{sec:min-ele}

In this section, we focus on the $\minknapsack$ problem, where the objective function is the minimum weight and the constraint is a knapsack constraint.  Specifically, letting $\F = \left \{ S \subseteq \universe \colon \cost{S} \le B \right \}$, the $\minknapsack$ problem can be written as:
\begin{equation}\label{eq:opt-min}
\min_{\pi \in \adm{\F}}{\ex{\min_{e \in \spi}{X_e}}}.
\end{equation}
Our main result is Theorem~\ref{thm:minelement}, restated below for convenience.
\RESTATEthmminele*

By the reduction to threshold problems in Lemma~\ref{lem:reduction}, to prove Theorem~\ref{thm:minelement}, it suffices to give a $(1,2)$-approximation to the $\minknapsack$ threshold problem:
\begin{equation}\label{eq:min-knapsack-threshold}
\min_{\pi \in \adm{\F}}{\pr{\min_{e \in \spi}{X_e} > t}}.
\end{equation}
In the remainder of this section, we show that a non-adaptive algorithm for this threshold problem achieves the desired $(1,2)$ approximation ratio.
More specifically, we first show that for the $\minknapsack$ threshold problem, a non-adaptive feasible policy achieves the adaptive optimum, i.e., the adaptivity gap is $1$, or equivalently, we say that there is no adaptivity gap. We then give a $(1,2)$-approximation for the non-adaptive version of the threshold problem.

\subsection{No adaptivity gap}
\begin{lemma}\label{lem:threshold-gap}
The $\minknapsack$ threshold problem in \eqref{eq:min-knapsack-threshold} has an adaptivity gap of $1$ for any $t\in \intdomain$.
\end{lemma}
\begin{proof}
Consider an arbitrary $t\in \intdomain$.  We induct on the size of the universe $\universe$.  Clearly, the adaptivity gap is $1$ when $\universe$ consists of one element.

Assume that the adaptivity gap is $1$ when the universe consists of $n$ elements for some $n\ge 1$; i.e., for any input of the threshold problem \eqref{eq:min-knapsack-threshold} such that the universe consists of $n$ elements, an optimal non-adaptive policy achieves the same objective value as an optimal adaptive policy.  Now consider any input such that the universe $\universe$ has $n+1$ elements and let $\policy^*$ be an optimal adaptive policy for it.  We will construct a non-adaptive policy, represented by a set $\set\subseteq\universe$, such that
\begin{equation}\label{eq:find-non-adaptive}
\p{\min_{e\in\set}X_e>t}=\p{\min_{e\in\set^{(\policy^*)}}X_e>t}.
\end{equation}

Without loss of generality, we can assume that $X_1$ is the first element probed by the optimal adaptive policy $\policy^*$ since the first probing decision does not depend on realizations of the random variables.  After $X_1$ is probed, we consider the threshold problem whose input consists of a universe $\universe'=\universe\setminus\{1\}$ and a budget $\budget'=\budget-c_1$.  Since $|\universe'|=n$, we know that this problem has an adaptivity gap of $1$.  Let $\set'\subseteq\universe'$ be the set chosen by an optimal non-adaptive policy for this problem.  

We claim that $\set=\{1\}\cup \set'$ satisfies \eqref{eq:find-non-adaptive}.  To see this, let $\policy^*_x$ for $x\in \intdomain$ be the subsequent policy of $\policy^*$ after seeing $X_1=x$, and let $\set^{(\pi^*_x)}\subseteq \universe'$ be the set chosen by $\policy^*_x$.  Then
\begin{align}
\p{\min_{e\in\set^{(\policy^*)}}X_e>t}&=\sum_{x=t+1}^{m}\p{X_1=x}\p{\min_{e\in\set^{(\policy^*)}}X_e>t\;\middle|\; X_1=x}\nonumber\\
&=\sum_{x=t+1}^{m}\p{X_1=x}\p{\min_{e\in\set^{(\policy^*_x)}}X_e>t}\label{eq:indpt}\\
&\ge \sum_{x=t+1}^{m}\p{X_1=x}\p{\min_{e\in\set'}X_e>t}\label{eq:induction}\\
&=\p{X_1>t}\p{\min_{e\in\set'}X_e>t}\nonumber\\
&=\p{\min_{e\in\set}X_e>t},\nonumber
\end{align}
where \eqref{eq:indpt} is due to the property of the minimum value, and \eqref{eq:induction} follows from the induction assumption that the non-adaptive choice of $\set'$ achieves optimality.  Since $\policy^*$ is an (adaptive) optimal policy, we know that this inequality is in fact an equality, which completes the proof.
\end{proof}

\subsection{Reduction to knapsack}
By virtue of Lemma~\ref{lem:threshold-gap}, to solve the threshold problem \eqref{eq:min-knapsack-threshold}, it suffices to solve its non-adaptive counterpart:
\begin{equation}\label{eq:opt-threshold-extreme-non}
\min_{\set \in \F}{ \pr{\min_{e\in\set}{X_e}>t}}
\end{equation}
Note that
\begin{equation*}
\pr{\min_{e\in\set}X_e>t}=\prod_{e\in\set}\pr{X_e>t}.
\end{equation*}
Taking a $\log$ of this probability, we rewrite \eqref{eq:opt-threshold-extreme-non} in the following equivalent form:
\begin{equation}\label{eq:opt-threshold-extreme-non-log}
\max_{\set \in \F}{ \sum_{e\in\set}\bigl(-\log\pr{X_e >t}\bigr)}
\end{equation}
This is a knapsack problem where the reward of each element $e$ is
$-\log\p{X_e>t}$.  Therefore, a $(1,2)$-approximation is given by
greedily adding elements in decreasing order of
$\frac{-\log\p{X_e>t}}{c_e}$ until the first time the total cost
exceeds the budget.  This completes the proof of Theorem~\ref{thm:minelement}.

\section{\texorpdfstring{The $\mink$-$\knapsack$ Problem}{The MinK-Knapsack Problem}}\label{sec:mink-knap}

In this section, we give the proof of Theorem~\ref{thm:mink-knap} alluded to in \S~\ref{sec:techniques}. We restate that theorem now, for the reader's convenience.

\RESTATEthmminkknap*

\newcommand{\idx}{j}

We begin with a definition.
For $i \in [k]$, if $\abs{S} \geq k$, let $y_i(S)$ be the $i$-th smallest weight in $S$.
With this, we can rewrite $f(S)$ as $\sum_{i=1}^k{y_i(S)}$. 
Noting that the functions $y_i(S)$ are monotonic in $i$, non-increasing in $S$, and take values in $\intdomain$,
we apply Corollary~\ref{coro:sum-of-k}, reducing the $\mink$-$\knapsack$ problem to the $y_i$-$\knapsack$ threshold problem. 
Let $\binset = \bin(\universe, B, i, t)$ be the output of the binning procedure in Algorithm~\ref{alg:binning-procedure} and let $$G = \greedyalg\left(\left\{e\in\universe\colon c_e\le \frac{B}{i}\right\}, B, i, t\right)$$ be the output of the extended greedy algorithm in Algorithm~\ref{alg:extended-greedy}.  
To complete the proof, we show that the non-adaptive policy $G\cup \binset$ is a $\minkknapthresh$-approximation for the $y_i$-$\knapsack$ problem. Before that though, we give an equivalent formulation of the threshold problem.
\begin{definition}
For a fixed threshold $t$, call an element $e$ \textbf{below-threshold} if $X_e \leq t$. Define $\trank{S}$ as the number of below-threshold elements contained in $S$, i.e.,
\begin{equation*}
    \trank{S} = \abs{\left \{ e\in S : X_e \leq t \right \}}.
\end{equation*}
\end{definition}
It follows from definitions that 
\begin{equation*}
\text{$y_i(S)\le t$ if and only if $\trank{S}\ge i$},
\end{equation*}
since both conditions imply and are implied by the presence of $i$ below-threshold elements in $S$.
Using the rank-based condition, the $y_i$-$\knapsack$ threshold problem
\begin{equation}
    \min_{\pi \in \adm{\F}}{\pr{y_i\left(\spi\right) > t}}.
\end{equation}
is equivalent to the following problem that we refer to as the \emph{$i$-th rank problem}
\begin{equation}\label{eq:rank}
\max_{\pi \in \adm{\F}}{\pr{\trank{\spi} \ge i }}.
\end{equation}
Note that the $i$-th rank problem is an instance of the $i$-heads problem defined in \S~\ref{sec:intro-mink} with heads-probability for element $e$ set to be $\pr{X_e\le t}$.
To complete our proof, it suffices to show that, for any $i \in [k]$ and $\pi \in \adm{\F}$,
\begin{equation}\label{eq:mink-value}
    \pr{\trank{G \cup \binset} \geq i} \geq \pr{\trank{\spi} \geq i}.
\end{equation}
and that
\begin{equation}{\label{eq:mink-cost}}
    \cost{G \cup \binset} \leq O(\log{k})B.
\end{equation}
We refer to these as the cost inequality and value inequality, respectively.

To see the cost inequality \eqref{eq:mink-cost}, note that for the $C_j$ in the binning procedure in Algorithm~\ref{alg:binning-procedure},
\begin{equation*}
    C_j \leq 2^j \cdot \frac{B}{2^{j-1}} = 2B
\end{equation*} and 
\begin{equation*}
    \cost{G} \leq B + (i+1)\cdot \frac{B}{i} \leq 3B.
\end{equation*}
Since $\binset = \cup_{j=1}^{\ceiling{\log{i}}}{C_j}$, it follows that
\begin{equation*}
    \cost{G \cup \binset} \leq 3B + \ceiling{\log{i}}\cdot 2B \leq O(\log{k})B,
\end{equation*}
where the last inequality comes from the fact that $i \leq k$.

The remainder of this section is devoted to proving the value inequality \eqref{eq:mink-value}. To do this, we show the following two lemmas.



\begin{restatable}{lemma}{RESTATElembinningprocedure}
\label{lem:binning-procedure-for-rank}
For the $i$-th rank problem with knapsack constraint $\F$ in \eqref{eq:rank}, suppose that there is a set of policies $\{\pi_{\ell}\colon \ell \in [i]\}$ such that for any $\ell\in[i]$ and any $\widehat{\pi} \in\adm{\F}$ whose $\ssup{\widehat{\pi}}$ only consists of elements from $\bigl\{e\in\universe\colon c_e\le \frac{B}{i}\bigr\}$,
\begin{equation*}
\pr{ \trank{\ssup{\pi_\ell}} \geq \ell} \geq \pr{\trank{\ssup{\widehat{\pi}}} \geq \ell}.
\end{equation*}
Let $\binset = \bin(\universe, B, i, t)$ be the output of the binning procedure.  If one probes $\binset$ and then executes $\pi_{\ell^*}$ with $\ell^* =\max\{i - \trank{\binset},1\}$, then the final selection set $\binset\cup \ssup{\pi_{\ell^*}}$ obtains an objective value in the $i$-th rank problem that is as good as the optimum, i.e., for any policy $\pi\in \adm{\F}$,
\begin{equation*}
    \pr{\trank{\binset \cup \ssup{\pi_{\ell^*}}} \geq i} \geq \pr{\trank{\ssup{\pi}} \geq i}.
\end{equation*}
\end{restatable}

\begin{restatable}{lemma}{RESTATElemextgreedy}
\label{lem:extgreedy-for-rank}
Consider the sequence of sets $\{G_{\ell}\colon \ell\in[i]\}$ defined by $G_{\ell}=\greedyalg\left(\{e\in\universe\colon c_e\le \frac{B}{i}\}, B, \ell, t\right)$.
Then for any $\ell\in[i]$ and any $\widehat{\pi} \in\adm{\F}$ whose $\ssup{\widehat{\pi}}$ only consists of elements from $\bigl\{e\in\universe\colon c_e\le \frac{B}{i}\bigr\}$,
\begin{equation*}
\pr{ \trank{G_{\ell}} \geq \ell} \geq \pr{\trank{\ssup{\widehat{\pi}}} \geq \ell}.
\end{equation*}
\end{restatable}

\newcommand{\low}{\textrm{low}}
\newcommand{\high}{\textrm{high}}

\begin{algorithm}
\caption{$\bin(\universe, B, i, t)$}
\label{alg:binning-procedure}
\begin{algorithmic}[1]
\State Initialization: $z \gets \ceiling{ \log{i} }$; $\binset \gets \emptyset$
\For{ $j$ from $1$ to $z$}
    \Comment{\textbf{Construct the cost bucket}}
    \State $\low \gets \max{\left \{\frac{B}{2^j}, \frac{B}{i} \right \}}$
    \State $\high \gets \frac{B}{2^{j-1}}$
    \State $\bucket_j \gets \left \{ e \in \universe \colon \low < c_e \leq \high \right \}$
    \State $C_j \gets \emptyset$
    \Comment{\textbf{Perform greedy selection}}
    \While{$\abs{C_j} < 2^j$ and $\bucket_j \neq \emptyset$}
        \State $ \ell \gets  \argmax_{e \in \bucket_j}{\pr{X_e \leq t}}$
        \State $\bucket_j \gets \bucket_j - \brk{\ell}$
        \State $C_j \gets C_j \cup \brk{\ell}$
    \EndWhile
    \State $\binset\gets \binset\cup C_j$
\EndFor
\State \Return $\binset$.
\end{algorithmic}
\end{algorithm}

\begin{algorithm}
\caption{$\greedyalg (\universe', B', i', t)$}
\label{alg:extended-greedy}
\begin{algorithmic}[1]
\State Initialization: $\delta \gets \max_{e \in \universe'}{c_e}$; $G \gets \emptyset$; $\pool \gets \universe'$

\For{ $ e \in \universe'$:}
\Comment{\textbf{Compute density of each element}}
    \State $\density \gets \frac{-\log{\pr{X_e > t}}}{c_e}$
\EndFor

\While{$\cost{G} < B' + i'\delta$}
\Comment{\textbf{Perform greedy selection}}
    \State $\ell \gets \argmax_{e \in \pool}{\density}$
    \State $G \gets G \cup \brk{\ell}$
    \State $\pool \gets \pool - \brk{\ell}$
\EndWhile
\State \Return $G$
\end{algorithmic}
\end{algorithm}

Combining Lemmas~\ref{lem:binning-procedure-for-rank} and \ref{lem:extgreedy-for-rank} completes the proof of the value inequality \eqref{eq:mink-value} once we notice that the $\{G_{\ell}\colon \ell\in[i]\}$ in Lemma~\ref{lem:extgreedy-for-rank} serves as the $\{\pi_{\ell}\colon \ell \in [i]\}$ in Lemma~\ref{lem:binning-procedure-for-rank} and $G_{\ell}\subseteq G$ for all $\ell\in[i]$, where one should recall that $G = \greedyalg(\{e\in\universe\colon c_e\le \frac{B}{i}\}, B, i, t)$. Thus, we now need only to prove the lemmas.

\subsection{Proof of Lemma~\ref{lem:binning-procedure-for-rank} (Analysis of $\bin$)}
We prove this lemma via two claims. First, in Claim~\ref{clm:binning-claim}, we show that the rank of the selection set $\spi$ under any policy $\pi$ is stochastically smaller than the rank of $\spi_{\leb{i}}\cup C$ where $\spi_{\leb{i}} \triangleq \spi \cap \{ e \in \universe : c_e \leq \frac{B}{i} \}$. In other words, we can replace the high-cost elements of $\spi$ with $\binset$ and \emph{strictly} improve its performance.
Second, in Claim~\ref{clm:beating-claim} we argue that, for any policy $\pi$, given knowledge of the set $\binset$'s rank, we can choose a policy $\pi'$ which selects only elements from $\{ e \in \universe : c_e \leq \frac{B}{i}\}$ whose set in expectation outperforms $\spi_{\leb{i}}$.
Note that the set $\spi_{\leb{i}}$ is a random set whose composition may depend on the weights of elements with costs larger than $\frac{B}{i}$.
Formally, we show the following claims.
\begin{claim}\label{clm:binning-claim}
For any selection set $\spi$,
\begin{equation*}
    \pr{\trank{\spi} \geq \target} \leq \pr{\trank{\spi_{\leq \frac{B}{i}}\cup \binset} \geq \target}
\end{equation*}    
\end{claim}

\begin{claim}\label{clm:beating-claim}
Let $\F' = \left \{ S \in \F: c_e \leq \frac{B}{i}, \forall e \in S  \right\}$ be the constraint family $\F$ but with all the high-cost elements removed. For any $\pi \in \adm{\F}$ and $\ell \in [\target]$, there exists a $\pi_\ell' \in \adm{\F'}$ such that
\begin{equation*}
    \pr{\trank{\spi_{\leq \frac{B}{i}} \cup \binset} \geq i \given \trank{\binset} = \target - \ell}
    \leq \pr{\trank{\ssup{\pi_\ell'}} \geq \ell}
\end{equation*}
\end{claim}
From here, the proof of Lemma~\ref{lem:binning-procedure-for-rank} flows quite directly:
\begin{align}
        \pr{\trank{\spi} \geq \target} &\leq \pr{\trank{\spi_{\leq \frac{B}{i}}\cup \binset} \geq \target}\label{eq:binp1}\\ 
        &= \sum_{\ell \in [\target]}{\pr{\trank{\binset} = \target - \ell}\pr{\trank{\spibi \cup \binset} \geq i \given \trank{\binset} = \target - \ell}}\label{eq:binp2}\\ 
        &\leq \sum_{\ell \in [\target]}{\pr{\trank{\binset} = \target - \ell}\pr{\trank{\ssup{\pi_\ell'}} \geq \ell}}\label{eq:binp3}\\ 
        &\leq \sum_{\ell \in [\target]}{\pr{\trank{\binset} = \target - \ell}\pr{\trank{\ssup{\pi_\ell}} \geq \ell}}\label{eq:binp4}\\\label{eq:bin-assumption}
        &= \pr{\trank{\binset \cup \ssup{\pi_{\target - \trank{\binset}}}} \geq \target},
\end{align}
where \eqref{eq:binp1} follows from Claim~\ref{clm:binning-claim}, \eqref{eq:binp3} follows from Claim~\ref{clm:beating-claim}, and \eqref{eq:binp4} follows from Lemma~\ref{lem:binning-procedure-for-rank}'s main assumption on $\{\pi_{\ell}\colon \ell\in[i]\}$. This completes our proof of Lemma~\ref{lem:binning-procedure-for-rank}, contingent on the claims.
\subsubsection{Proof of Claim~\ref{clm:binning-claim}}
To show that 
\begin{equation*}
        \pr{\trank{\spi} \geq \target} \leq \pr{\trank{\spi_{\leq \frac{B}{i}}\cup \binset} \geq \target},
\end{equation*}
we make a simple coupling argument. Consider the following procedure: Execute $\pi$ but make the modification that, every time the policy $\pi$ would probe an element $e$ of cost $>\frac{B}{\target}$, instead have $\pi$ do one of the following three things, where we, without loss of generality, let $j$ be such that $e \in \bucket_j$:
\begin{enumerate}
    \item  If $e \not \in C_j$, then we instead make $\pi$ probe the unprobed element in $C_j$ of smallest below-threshold probability. In other words, $\pi$ probes
    \begin{equation*}
        \argmin_{v \in C_j - \spi}{\pr{X_v \leq t}},
    \end{equation*}
    where $\spi$ is the policy $\pi$'s current selection set (instead of its final selection set, which is the usual definition).
    \item If $e \in C_j$ and hasn't been selected yet, then we allow it to proceed as usual, i.e., we allow $\pi$ to probe $e$. 
    \item If $e \in C_j$ but has already been probed (because of the first line in this list), we direct $\pi$ to probe the element $v \in C_j - \spi$ of immediately higher $\pr{X_v \leq t}$. Note that this is always possible, since $\pi$ probes at most $2^j$ elements from $\bucket_j$, and we choose elements with higher $\pr{X_v \leq t}$'s first.
\end{enumerate}
If this procedure is applied, every probed element of cost $> \frac{B}{i}$ must lie in $\binset$.  Note also that, in all cases, the element we choose must have a below-threshold probability at least as big as the element $e$ being replaced, since any $C_j$ contains the $2^j$ elements from $\bucket_j$ with the highest below-threshold probability. For any optimal policy, it follows that this would result in a strict increase in success probability. But every element in our procedure must lie in $\binset$, and probing the rest of $\binset$ can only improve things, which implies
\begin{equation*}
        \pr{\trank{\spi} \geq \target} \leq \pr{\trank{\spi_{\leq \frac{B}{\target}}\cup \binset} \geq \target},
\end{equation*}
as desired.

\subsubsection{Proof of Claim~\ref{clm:beating-claim}}
To see that
\begin{equation*}
    \pr{\trank{\spi_{\leq \frac{B}{i}}\cup C} \geq i \given \trank{\binset} = \target - \ell}
    \leq \pr{\trank{\ssup{\pi_\ell'}} \geq \ell}
\end{equation*}
it suffices to consider the following thought experiment: Give that a policy free access to $C$, and full knowledge of $\left \{X_e : e \in \binset \right \}$, but restrict it to selecting new elements of cost $> \frac{B}{i}$; how would that policy maximize $\pr{\text{selection set reaches rank $i$}}$? One approach is to
run $\pi$ using the replacement procedure from the proof of the previous claim; whenever $\pi$ would probe an element $e$ of cost $> \frac{B}{i}$, we can just pretend it looked at $X_e$ while using the weight of its replacement element. In other words, we can \emph{simulate} having access to the full set of high-cost elements, while \emph{actually} only having access to $\binset$.

Call a policy low-cost if it does not choose elements of cost $> \frac{B}{i}$. A better way is to take the remaining rank required, $\ell = i - \trank{\binset}$, and find a low-cost policy $\pi'$ which maximizes $\pr{\trank{\ssup{\pi'}} \geq \ell \given \cap_{e \in \binset} {\left \{X_e = x_e \right \}}}$. Clearly though, from the independence of weights, the performance of any valid policy which probes no elements in $\binset$ must be independent of the outcomes in $\binset$, i.e.,
\begin{equation*}
    \pr{\trank{\ssup{\pi'}} \geq \ell \given \cap_{e \in \binset} \left \{X_e = x_e \right \}} = \pr{\trank{\ssup{\pi'}} \geq \ell \}}.
\end{equation*}
Take $L$ to be the set of $(x_e)_{e\in \binset}$ values where $\trank{\binset} = i - \ell$. Since by assumption $\pi_\ell'$ is \emph{better} than $\pi'$, it follows from the above that
\begin{align*}
     \pr{\trank{\binset} = \target - \ell} 
     &\pr{\trank{\spi_{\leq \frac{B}{i}}\cup C} \geq i \given \trank{\binset} = \target - \ell}\\
    &=  \pr{\trank{\binset} = \target - \ell} 
     \pr{\trank{\spi_{\leq \frac{B}{i}}} \geq \ell \given \trank{\binset} = \target - \ell}\\
    &= \sum_{ (x_e)_{e \in \binset} \in L }
        {\pr{\cap_{e \in \binset} {\left \{X_e = x_e \right\}}}
        \pr{\trank{\spi_{\leq \frac{B}{i}}} \geq \ell \given  \cap_{e \in \binset} \left \{X_e = x_e \right \}}
        }\\
     &\leq \sum_{ (x_e)_{e \in \binset} \in L }
        {\pr{\cap_{e \in \binset} {\left \{X_e = x_e \right\}}}
        \pr{\trank{\ssup{\pi'}} \geq \ell \given  \cap_{e \in \binset} \left \{X_e = x_e \right \}}
        }\\
    &= \sum_{ (x_e)_{e \in \binset} \in  L}
        {\pr{\cap_{e \in \binset} {\left \{X_e = x_e \right\}}}
        \pr{\trank{\ssup{\pi'}} \geq \ell }
        }\\
    &= \left[ \sum_{ (x_e)_{e \in \binset} \in L}
        {\pr{\cap_{e \in \binset} {\left \{X_e = x_e \right\}}}} \right]
        \pr{\trank{\ssup{\pi'}} \geq \ell 
        }\\
    &= \pr{\trank{\binset} = \target - \ell} \pr{\trank{\ssup{\pi'}} \geq \ell 
        }\\
    &\leq \pr{\trank{\binset} = \target - \ell} \pr{\trank{\ssup{\pi_{\ell}'}} \geq \ell 
        },
\end{align*}
from which the claim 
\begin{equation*}
    \pr{\trank{\spi_{\leq \frac{B}{i}}\cup C} \geq i \given \trank{\binset} = \target - \ell}
    \leq \pr{\trank{\ssup{\pi_\ell'}} \geq \ell}
\end{equation*}
follows easily.



\subsection{Proof of Lemma~\ref{lem:extgreedy-for-rank} (Analysis of \textsc{ExtGreedy})}
We now prove to prove the remaining lemma, which concerns $\greedyalg$, Algorithm~\ref{alg:extended-greedy}.


Throughout this proof, we assume, for brevity's sake, that $\universe$ contains no elements of cost $> \frac{B}{i}$ and that the constraint $\F$ is a knapsack constraint using the original budget $B$ and this limited-cost universe $\universe$. This in no way affects the proofs here, since $G$ in the algorithm proper is drawn from a limited cost universe and the policy it is competing against, $\widehat{\pi}$, must be admissible in this low-cost setting.

We proceed in four steps. Recall that we say a policy $\pi$ succeeds if $\trank{\spi} \geq \ell$. First, we provide a characterization of $\pr{\text{$\pi$ succeeds}}$ based on a structural decomposition of $\pi$'s decision tree. Second, we use that characterization to derive a sufficient condition on two policies $\pi$ and $\pi'$ which implies that $\pr{\text{$\pi'$ succeeds}} \geq \pr{\text{$\pi$ succeeds}}$. In the last two steps, we prove that this sufficient condition holds  
for $G_\ell$ and $\pi$.

\subsubsection{Characterization of success probability}\label{sec:mink-succ-char}
In the first step of our proof, we characterize the success probability of policies for the $\ell$-th rank problem. First, we decompose the structure of a policy $\pi$, then we describe how to use that decomposition to represent the probability that $\pi$ succeeds.

\paragraph{Policies and decision trees.} We assume without loss of generality that a policy is constructed as a decision tree, where nodes correspond to elements probed, and edges correspond to the particular value of an element's weight upon probing. Because we are concerned only with $\trank{S}$, and element-weights are independent of each other, one can, without lowering a policy's probability of success, separate  the $(m+1)$-many possible outcomes into two particular meta-outcomes: whether the element $e$ probed has $X_e \leq t$, or $X_e > t$. We also associate a particular directionality to outcomes: the root is the first element probed, the left child of the root is the element probed if the first element probed is below-threshold, and so on.

\begin{definition}[The Left Chain, $\Lpi$]
     Define the ``left chain'' of a decision tree to be the sequence of elements probed when every previous element probed is above-threshold. Let $\lpi_j$ be the $j$-th element in this sequence. Then, the left chain of a policy $\pi$ is simply the sequence
\begin{equation}
    \Lpi \triangleq \left(\lpi_0,\lpi_1, \dots \right).
\end{equation}
Further, let $\Lpi_j \triangleq \left(\lpi_0, \lpi_1, \dots, \lpi_j \right)$ be the left chain of $\pi$, truncated at the $j$-th element.
\end{definition}

\begin{figure}
    \begin{center}
        \scalebox{0.6}{
            \begin{tikzpicture}[
        triangle/.style = {fill=blue!20, regular polygon, regular polygon sides=3, scale=1.2, draw, thick},
        circ/.style = {fill=red!20, circle, inner sep=15, draw, thick}
]
    \node[triangle, thick] at (3,-3.5) (aT) {$T^{(\pi)}_0$};
    \node[triangle, thick] at (0,-6) (bT) {$T^{(\pi)}_1$};
    \node[triangle, thick] at (-2,-9) (cT) {$T^{(\pi)}_2$};

    \node at (.25,0.375) (pre) {};
    \node[circ] at (0, 0) (a) {};
    \node[circ] at (-2, -3) (b) {};
    \node[circ] at (-4, -6) (c) {};
    \node[circ] at (-6, -9) (d) {};
    \node at (-6.25, -9.375) (post) {};
    
    \node at (-4.25, -6.375) (post2) {};
    
    \node[scale=1.2] at (0,0) (apt) {$\lpi_0$};
    \node[scale=1.2] at (-2,-3) (bpt) {$\lpi_1$};
    \node[scale=1.2] at (-4,-6) (cpt) {$\lpi_2$};
    \node[scale=1.2] at (-6,-9) (dpt) {$\lpi_{\abs{\Lpi}}$};

    
    
    \draw[->, ultra thick] (a) -- (aT.north) node[midway, fill=white] {$\weightfig{\lpi_0}\leq t$};
    \draw[->, ultra thick] (a) -- (b) node[midway, fill=white] {$\weightfig{\lpi_0} > t$  };
    
    \draw[->, ultra thick] (b) -- (bT.north);
    \draw[->, ultra thick] (b) -- (c);
    
    \draw[->, ultra thick] (c) -- (cT.north);
    \draw[->, ultra thick] (c) -- (d) node[midway,fill=white,scale=2] {$\ldots$};
    
    \draw[decoration={brace, mirror, raise=30pt, amplitude=15pt},decorate,  thick]
  (pre) -- node[above=32pt, left=30pt, scale=1.2] {$\Lpi_2$} (post2);
  
    \draw[decoration={brace, mirror, raise=50pt, amplitude=40pt},decorate,  thick]
  (pre) -- node[above=60pt, left=70pt, scale=1.2] {$\Lpi$} (post);

    \end{tikzpicture}
        }
    \caption{A visualization of the structural decomposition. Upon probing an element $\lpi_j$, if $X_{\lpi_j} \leq t$, then the policy $\pi$ proceeds to the right child of the node ($\lpi_j$); otherwise, $\pi$ proceeds to the left child. The left child of the node ($\lpi_j$) is $\lpi_{j+1}$, the next node in the left chain $\Lpi$. The right child is the exit tree $T^{(\pi)}_j$.}
    \label{fig:left-chain}
    \end{center}
\end{figure}
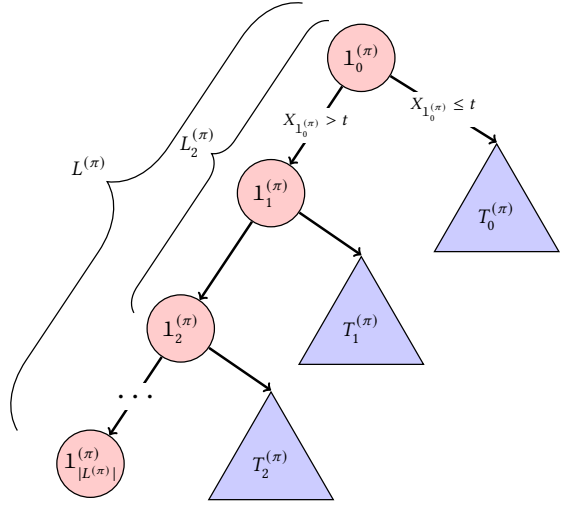

\begin{definition}[Exit trees, $\Tpi_j$] 
We say the policy $\pi$ ``exits the left chain'' at $j$ if $\lpi_j$ is the first below-threshold element probed. It follows that, for every $j$, one can talk about the decision sub-tree corresponding to ``exiting'' at $j$; we call this sub-tree $\Tpi_j$.  
\end{definition}
Together with $\Lpi$, these sub-trees give a natural decomposition of the policy $\pi$; we visualize this decomposition in Figure~\ref{fig:left-chain}.

\paragraph{Non-adaptive policies.} One can also apply this decomposition to non-adaptive policies (deterministic sets) $G$. In fact, one special characteristic of non-adaptive policies is that there is no fixed left chain: one can probe the deterministic set $G$ in any order, without changing the probability that $\trank{G} \geq \targetnew$. In other words, one can define $\LG$ to be an arbitrary permutation of $G$ without affecting the probability that $G$ succeeds. Note that, by definition, the ``exit trees'' of $G$ are simply the policies which probe whatever elements in $G$ remain unprobed. In other words, 
\begin{equation}
    \TG_j = G - \LG_j.
\end{equation}

We next characterize the success probability $\pr{\trank{\spi} \geq \ell}$ under a policy $\pi$ based on the structural decomposition.
\begin{definition}[The Exit Value, $\Rpi_j$]\label{def:exit-value}
For a policy $\pi$, let the exit value $\Rpi_j$ be the probability of the policy $\pi$'s success given it has exited the left chain after probing the element $\lpi_j$, i.e., 
\begin{equation*}
    \Rpi_j \triangleq \Pr \left( \trank{\spi} \geq \targetnew  \given \trank{\Lpi_{j-1}} = 0, \lpi_j \text{ is below-threshold}\right).
\end{equation*}
\end{definition}
For each possible exit $j$, let $\F'$ be a knapsack constraint with universe $\universe' = \universe - \Lpi_j$ and budget $B' = B - \cost{\Lpi_j}$. By element-weight independence, the behavior of policy $\pi$ after leaving the left chain at $j$ can be treated, without loss of generality, as the behavior of a some new policy $\pi'$ with constraint $\F'$. In other words,
\begin{align*}
    \Rpi_j = \pr{\trank{\ssup{\pi}} \geq \ell \given \text{$\pi$ exits at $j$}} = \Pr \left( \trank{S^{(\pi')}} \geq \targetnew - 1 \right).
\end{align*}

\paragraph{Basic characterization of a policy $\pi$.} Let $\ppi_j$ be the probability that the left-element $\lpi_j$ is below-threshold, and, for notational clarity, let $x$ be the length of the sequence $\Lpi$. 
Together with the left chain and exit values of a policy $\pi$ defined, we can characterize the success probability of a policy as
\begin{align*}\label{eq:minknap_pre}
    \succprob{\spi}{\targetnew} &= \ppi_0 \cdot \Rpi_0\\
    &\; + \left(1 - \ppi_0 \right)\ppi_1 \cdot \Rpi_{1}\\
    &\; +\, \dots \mathop{}\\
    &\; + \left[ \prod_{y=0}^{j - 1}{\left( 1 - \ppi_y \right)}\right] \ppi_j \cdot \Rpi_{j}\\
    &\; +\, \dots\\ 
    &\; + \left[ \prod_{y=0}^{x-1}{\left( 1 - \ppi_y \right)}\right] \ppi_x \cdot \Rpi_{x}
\end{align*}
We develop this decomposition further.
\begin{definition}[The width, $\width{S}$]
For a set $S$, define the width of $S$, denoted $\width{S}$, to be the probability that at least one element in $S$ is below-threshold.
\end{definition}
Note that $\width{S}$ can be written as
\begin{align*}
\width{S} = 1 - \prod_{e \in S}{\left(\pr{X_e > t} \right)}
= 1 - \exp\left(\sum_{e \in S}{\log\left({\pr{X_e > t}} \right)} \right).
\end{align*}
Let the reward of an element $e$ be $\reward{e} = - \log \left( \pr{X_e > t} \right)$, and the total reward of a set $S$ by extension be $\reward{S} 
= \sum_{e \in S}{\reward{e}}$. 
Then $\width{S}$ is an increasing function of $\reward{S}$ since
\begin{align*}
    \width{S} &= 1 - \exp\left( -\sum_{e \in S}{\reward{e}} \right)\\
            &\triangleq \widthsum{ \sum_{e \in S}{\reward{e}}},
\end{align*}
where $\widthsum{y} = 1 - \exp(-y)$ \emph{is} that increasing function.
Notice that, for any element $u \not \in S$,
\begin{equation}\label{eq:width_decomp}
    \left[ \prod_{e \in S}{\left(1 - p_e\right)} \right] p_u = \width{S \cup \brk{u}} - \width{S}.
\end{equation}

Making use of (\ref{eq:width_decomp}), we have that, for any policy $\pi$ with a left chain of length $x$,
\begin{equation}\label{eq:full_decomp}
\begin{split}
    \succprob{\spi}{\targetnew} &= \width{\Lpi_0} \cdot \Rpi_0\\ 
    &\; + \left[\width{\Lpi_1} - \width{\Lpi_0} \right]\cdot \Rpi_1 \\
    &\; +\, \dots \\
    &\; + \left[ \width{\Lpi_{j+1}} - \width{\Lpi_j} \right] \cdot \Rpi_{j+1} \\
    &\; +\, \dots\\
    &\; + \left[ \width{\Lpi_{x}} - \width{\Lpi_{x-1}} \right] \cdot \Rpi_{x}.
\end{split}
\end{equation}

\subsubsection{A sufficient condition: Rectangle covering}

\begin{figure}
    \begin{center}
    \scalebox{0.7}{
            \begin{tikzpicture}
        \filldraw [color=black!50, fill opacity= 0.65, ultra thick] (0,0) rectangle (7, 1.5);    
        \filldraw [color=blue!60, fill opacity= 0.65, ultra thick] (0,0) rectangle (6, 2);    
        \filldraw [color=red!80, fill opacity= 0.65, ultra thick] (0,0) rectangle (4, 3);
        
        \node[scale=2] at (8, 0.75) {$\ldots$};

        \node at (2,2.5) (firstreg) {};
        \node at (4,4) (firstarea) {$\width{\Lpi_0}\cdot \Rpi_0$};
        \draw[thick] (firstreg) -- (firstarea);
        
        \node at (5,1.75) (secreg) {};
        \node at (7.5,3.00) (secarea) {$\left( \width{\Lpi_1} - \width{\Lpi_0} \right) \cdot \Rpi_1$};
        \draw[ thick] (secreg) -- (secarea);
        
        \node at (6.5,1) (thirdreg) {};
        \node at (10,2.00) (thirdsec) {$\left( \width{\Lpi_2} - \width{\Lpi_1} \right) \cdot \Rpi_2$};
        \draw[thick] (thirdreg) -- (thirdsec);

        \draw[decoration={brace, mirror, raise=2pt},decorate]
  (0,0) -- node[below=2pt] {$\width{\Lpi_0}$} (4,0);

        \node at (0,-1.05) (a1) {};
        \node at (6, -1.05) (b1) {};
        \draw[dotted] (0,0) -- (a1);
        \draw[dotted] (6,0) -- (b1);
         \draw[decoration={brace, mirror, raise=0pt},decorate]
  (a1) -- node[below=2pt] {$\width{\Lpi_1}$} (b1);

        \node at (0,-2.1) (a2) {};
        \node at (7, -2.1) (b2) {};
        \draw[dotted] (0,0) -- (a2);
        \draw[dotted] (7,0) -- (b2);
         \draw[decoration={brace, mirror, raise=0pt},decorate]
  (a2) -- node[below=2pt] {$\width{\Lpi_2}$} (b2);
  
        \draw[->,thick] (0,0) -- (0,4);
        \draw [->, thick] (0,0) -- (10,0);
    \end{tikzpicture}
    }
    \end{center}
    \caption{A visualization of the success probability decomposition found in $\eqref{eq:full_decomp}$. The $j$-th rectangle has width $\width{\Lpi_j}$ and height $\Rpi_j$. The total area covered by these rectangles is the success probability $\pr{\trank{\spi}\ge \ell}$. Full details can be found in \S~\ref{sec:mink-succ-char}.}
    \label{fig:rect-decomp-tikz}
\end{figure}
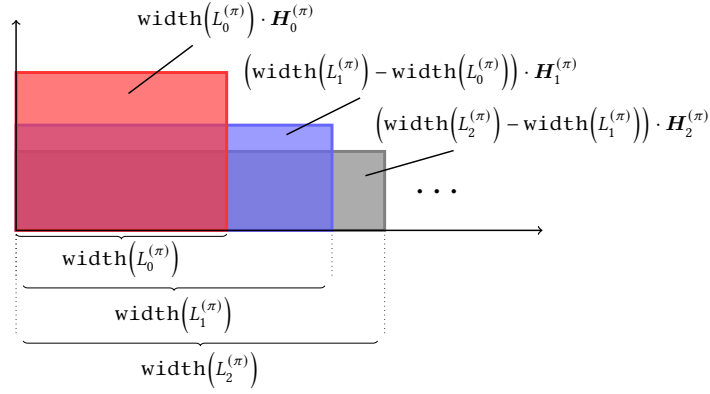

In the second step of our proof, we use the characterization just developed to derive a sufficient condition on two policies $\pi$ and $\pi'$ which implies that $\pr{\text{$\pi'$ succeeds}} \geq \pr{\text{$\pi$ succeeds}}$. We make a visual argument. One can characterize the summation in \eqref{eq:full_decomp} as the total area covered by a sequence of rectangles all nestled in the corner of the positive orthant, where the $j$-th rectangle in the sequence has height $\Rpi_j$ and width $\width{\Lpi_j}$; we visualize this in Figure~\ref{fig:rect-decomp-tikz}.

Likewise, for any \textit{non-adaptive} policy $G$, for \emph{any} left-chain ordering $\LG$ on the set $G$, one has analogously
\begin{align}
    \succprob{G}{\targetnew} &= \width{\LG_0}\succprob{G - \LG_0}{\targetnew-1}\\
    &\; +  \sum_{j=1}^{\abs{G}}
        {
            \left[ \width{\LG_j} - \width{\LG_{j-1}} \right]
            \cdot
            \succprob{G - \LG_j}{\targetnew-1}
        }, 
\end{align}
i.e., the previous decomposition of \eqref{eq:full_decomp}, except now with $\RG_j = \succprob{G - \LG_j}{\targetnew-1}$.

\paragraph{Main idea.} From a geometric standpoint then, to show that a non-adaptive policy $G$ succeeds more often than an optimal policy $\pi$, it suffices to show that the area covered by $G$ in such a diagram is larger than the area covered by $\pi$ in such a diagram. Further, it suffices to show that the each of $\pi$'s rectangles is \emph{completely covered} by at least one $G$'s rectangles, for some ordering of the elements on the left chain $\LG$. We shall do precisely this, with $G = \greedyalg(\universe, B, \targetnew, t)$. Formally, we will show the following lemma. 

\begin{lemma}[Rectangle-covering lemma]\label{lem:rect_covering}
Let $G = \greedyalg(\universe, B, \targetnew, t)$. Given an optimal policy $\pi$'s left chain $\Lpi$, one can construct a fixed ordering $\LG$ such that there exists an increasing sequence of indices $\left(\sigma(j) \right)_{j \in \left(0,1,\dots, \abs{\Lpi} \right)}$ satisfying for every $j$
\begin{equation}\label{eq:mink_width}
     \width{\LG_{\sigma(j)}}\geq \width{\Lpi_j}
\end{equation}
and 
\begin{equation}\label{eq:mink-rval}
    \RG_{\sigma(j)} \geq \Rpi_j.
\end{equation}
In other words, the $\sigma(j)$-th rectangle of $G$ \emph{completely covers} the $j$-th rectangle of $\pi$.
\end{lemma}
Note that the sequence of $\sigma(j)$'s in Lemma~\ref{lem:rect_covering} need not be consecutive. For example, the first rectangle of $G$ might completely cover the first 4 rectangles of $\pi$. In that case, one could set $\sigma(1)$, $\sigma(2)$, $\sigma(3)$, and $\sigma(4)$ all to be $1$. The $\sigma(\cdot)$ sequence is simply a formalism by which we can explicitly assign coverage of a particular $\pi$-rectangle to a particular $G$-rectangle.

\subsubsection{Proof of Lemma~\ref{lem:rect_covering}} \label{sec:rect_covering}
All that remains for proving Lemma~\ref{lem:extgreedy-for-rank} is to prove Lemma~\ref{lem:rect_covering}. 
We prove Lemma~\ref{lem:rect_covering} by induction on the target rank $\targetnew$. 

In the base case, $\targetnew = 1$, and whenever a policy exits its left chain, it has succeeded. In other words, for any policy $\pi$ and index $j$, the height $\Rpi_j = 1$. To prove the base case, since all rectangle heights are the same, it suffices to show the width inequality $\width{\LG} \geq \width{\Lpi}$; equivalently, we must show that $ \reward{\LG}\geq \reward{\Lpi}$. But this follows directly from our greedy-until-overflowing strategy for selecting $G$, proving the base case.

For the remainder of this section, we inductively assume the following inductive assumption.
\paragraph{Inductive assumption.} Given any universe $\universe'$, budget $B'$, target $\targetnew' < \targetnew$, and threshold $t$, for any policy $\pi' \in \adm{\F'}$, one can construct a fixed ordering $L^{(G')}$ of the set $G' = \greedyalg(\universe', B', \targetnew', t)$ such that there exists an increasing sequence of indices $\left(\beta(j) \right)_{j \in \left(0,1,\dots, \abs{L^{(\pi')}}\right)}$ satisfying
\begin{equation*}
    \width{L^{(G')}_{\beta(j)}} \geq \width{L^{(\pi')}_j} 
\end{equation*}
and 
\begin{equation*}
\Hsup{G'}_{\beta(j)} \geq  \Hsup{\pi'}_j.
\end{equation*}

\paragraph{Inductive step.}
For the remainder of our proof, we use our inductive assumption for lower-rank problems to construct a left chain ordering $\LG$ and assignment sequence $\sigma(\cdot)$ which satisfies these conditions for the $\ell$-th rank problem. Before we continue with our construction, we define a surrogate policy $\pitilde$ for the policy $\pi$ which simplifies our analysis. 


\paragraph{Construction of the surrogate policy $\pitilde$.}
\begin{definition}[The residual sets, $\Tstar_j$.]
Let 
\begin{equation*}
    \Tstar_j = \greedyalg(\universe - \Lpi_j, B - \cost{\Lpi_j}, \targetnew - 1, t)
\end{equation*} be the extended greedy set one would choose using the resources available after exiting the left chain at $j$ while executing $\pi$. As usual, we refer to both the non-adaptive policy which probes $\Tstar_j$ and the set $\Tstar_j$ by the same name.
\end{definition}

Given a policy $\pi$ and the residual sets $\Tstar_j$, we now construct a simpler-to-analyze policy $\pitilde$ such that 
\begin{equation*}
    \Pr(\text{$\pitilde$ succeeds}) \geq \Pr(\text{$\pi$ succeeds}).
\end{equation*}
\begin{claim}
    Recall the decomposition of $\pi$ into $\Lpi$ and sub-trees $\Tpi_j$. Likewise, define $\pitilde$ as the policy which has the same left chain as $\pi$, but replaces each sub-tree $\Tpi_j$ with the set $\Tstar_j$. Then,
    \begin{equation}
        \succprob{\spitilde}{\targetnew} \geq \succprob{\spi}{\targetnew}.
    \end{equation}
\end{claim}

\begin{proof}
We show first that $\pitilde$ is a valid policy. Let $\universe' = \universe - \Lpi_j$, let $B' = B - \cost{\Lpi_j}$, and let $\F'$ be a knapsack constraint formed with universe $\universe'$ with budget $B'$. Since the behavior of policy $\pi$ after leaving the left chain at $j$ is equivalent to the behavior of some new policy $\pi'$ with constraint $\F'$, and the set $G'= \greedyalg(\universe', B', \targetnew - 1, t)$ must be disjoint from $\Lpi_j$, it follows that $\pitilde$, which probes the left chain $\Lpi$ until we exit at $j$ then probes $G'$, is a perfectly valid policy; it probes no elements twice, and uses no foreknowledge of any weight outcomes.

We now show that 
\begin{equation*}
    \Pr(\text{$\pitilde$ succeeds}) \geq \Pr(\text{$\pi$ succeeds}).
\end{equation*}
We argue this from the success probability characterization of \eqref{eq:full_decomp}. Since $\pitilde$ and $\pi$ share the same left chain $\Lpi$, they must have the same width values $\width{\Lpi_j}$. Thus, it suffices to show that replacing $\pi'$ with $G'$ can only increase the height of $\pitilde$, i.e., that
\begin{equation*}
     \Hsup{\pitilde}_j \geq \Rpi_j.
\end{equation*}
To see this, note that, by the same justification that says Lemma~\ref{lem:rect_covering} (rectangle covering) implies Lemma~\ref{lem:extgreedy-for-rank} ($\greedyalg$ is better than $\pi$), our inductive assumption immediately implies that, since $\pi' \in \adm{\F'}$,
\begin{equation*}
    \succprob{G'}{\targetnew - 1} \geq \succprob{\ssup{\pi'}}{\targetnew-1}.
\end{equation*} 

Since, by the definition of $\pi'$, 
\begin{equation*}
\pr{\trank{\spi} \geq \ell \given \text{$\pi$ exits the left chain at $j$}} = \pr{\trank{\ssup{\pi'}} \geq \ell - 1},
\end{equation*}
we have that
\begin{align*}
    \Hsup{\pitilde}_j &= \succprob{G'}{\targetnew-1}\\
    &\geq \pr{\trank{\ssup{\pi'}} \geq \ell - 1}\\
    &= \pr{\trank{\spi} \geq \ell \given \text{$\pi$ exits the left chain at $j$}}\\
    &= \Rpi_j,
\end{align*}
as desired.
\end{proof}

\newcommand{\uhat}{\widehat{\universe}}
\newcommand{\bhat}{\widehat{B}}

\paragraph{Construction of $\LG$ and $\sigma(\cdot)$.}

We now proceed with our construction. We first make the following claim about the sets $\Tstar_j$.
\begin{claim}\label{clm:mink-inclusion}
    Let $\pi$ be any policy in $\adm{\F}$ and let $\Lpi$ be its left chain. Let \begin{equation*}
        \Tstar_j = \greedyalg \left(\universe - \Lpi_j, B - \cost{\Lpi_j}, \targetnew -1, t \right).
    \end{equation*} 
    Then, for all $j \in [\abs{\Lpi}]$
    \begin{equation}
        \Tstar_{j+1} \subseteq \Tstar_{j} \subseteq G.
    \end{equation}
\end{claim}
\begin{proof}
We first find a simpler sufficient condition to show. As usual, for a fixed index $j$, let $\universe' = \universe - \Lpi_j$ and let $B' = B - \cost{\Lpi_j}$. Now, let $\delta'$ be the max cost of any element in $\universe'$. By the definition of $\greedyalg$,
\begin{equation*}
    \greedyalg \left(\universe', B', \targetnew - 1, t \right) = \greedyalg \left( \universe', B' + (\targetnew - 1) \delta', 0, t \right). 
\end{equation*}
Furthermore, as the maximum cost of any element in $\universe$ can only decrease as elements are removed, $\delta' \leq \delta$, and, from a budget argument,
\begin{equation*}
    \greedyalg \left(\universe', B' + (\targetnew-1)\delta', 0, t \right) \subseteq     \greedyalg \left(\universe', B' + (\targetnew-1)\delta, 0, t \right).
\end{equation*}
As such, we have that
\begin{equation*}
    \Tstar_j = \greedyalg \left(\universe', B' + (\targetnew-1)\delta', 0, t \right) 
\end{equation*}
and, letting $e = \lpi_j$ be the $j$-th element in the left chain $\Lpi$,
\begin{equation*}
    \Tstar_{j+1} \subseteq \greedyalg \left(\universe'-\brk{e}, B' - c_e + (\targetnew-1)\delta', 0, t \right).
\end{equation*}
Thus, to show 
\begin{equation*}
\Tstar_{j+1} \subseteq \Tstar_{j} \subseteq G,
\end{equation*}
we prove a more general claim; that, for a general universe $\uhat$, budget $\bhat$, and element $e \in \widehat{\universe}$, 
\begin{equation*}
  \greedyalg \left(\uhat - \brk{e}, \bhat - c_e, 0, t \right) \subseteq    \greedyalg \left(\uhat, \bhat, 0, t \right).
\end{equation*}


For brevity, call the former set $T'$ and the latter $T$.
Without loss of generality, number the elements of $\uhat$ from $1$ to $\abs{\uhat}$ in descending order of reward density, maintaining that higher density elements have lower element number. 
Let $\tau$ be the smallest index for which 
\begin{equation*}
\sum_{x=1}^{\tau}{c_x} \geq \bhat.
\end{equation*}
Then, by definition,  the set $T =  \{1, 2, \dots, \tau\}$. Likewise, if $\tau'$ is the smallest index for which 
\begin{equation*}
    \sum_{x=1, x\neq e}^{\tau'}{c_x} \geq \bhat - c_e,
\end{equation*} then $T' = \{ 1, 2, \dots, \tau'\} - \brk{e}$. Thus, to show $T' \subseteq T$, it suffices to show that $\tau' \leq \tau$.

There are two cases. If the removed element $e \leq \tau$, then the selection process for $T'$ ends by $\tau$ at the latest, since
\begin{equation*}
    \sum_{x=1}^{\tau}{c_x} \geq \bhat
\end{equation*}
implies that
\begin{equation*}
    \sum_{x=1, x\neq e}^{\tau}{c_x} = \sum_{x=1}^{\tau}{c_x} - c_e \geq \bhat - c_e,
\end{equation*}
and $\tau'$ is the smallest index which satisfies this condition. Hence, $\tau' \leq \tau$.

In the other case, the removed element $e > \tau$, and
\begin{equation*}
\sum_{x=1}^{\tau}{c_x} = \sum_{x=1, x \neq e}^{\tau}{c_x},
\end{equation*}
meaning
\begin{equation}\label{eq:taueq2}
\sum_{x=1, x \neq e}^{\tau}{c_x} \geq \bhat \geq \bhat - c_e.
\end{equation}
Since $\tau'$ is the smallest index for which \eqref{eq:taueq2} is true, $\tau' \leq \tau$, proving the claim.

\end{proof}

This claim almost directly completes the construction. First, it induces a total order on $G$, where the order of an element $e \in G$ is the smallest index $j$ for which $e \not \in \Tstar_j$, i.e., all the elements of $G - \Tstar_0$ are of order $0$, all the elements of $\Tstar_0 - \Tstar_1$ are of order $1$, and so on. We can then take $\LG$ to be any permutation of $G$ which proceeds in non-decreasing order. Second, it allows us to define $\sigma(j)$ as the index $r$ for which
\begin{equation*}
    G - \LG_r = \Tstar_j;
\end{equation*}
in other words, we make $\sigma(j) = \abs{ \left\{e \in G: \text{$e$'s order} \leq j \right\}} - 1$. This completes the desired construction.

\paragraph{Verifying the construction.}
Having given a construction for $\LG$ and $\sigma(\cdot)$, in this final part of our proof, we show that the given construction satisfies Lemma~\ref{lem:rect_covering}'s width and height conditions. 
The height condition is satisfied by construction, since, if $\pitilde$ exits the left chain at $j$, then it probes $\Tstar_j$, the same set that $G$ probes upon exiting its left chain at $\sigma(j)$. 
All that remains is to show the width condition
\begin{equation*}
    \width{\LG_{\sigma(j)}} = \width{G - \Tstar_j} \geq \width{\Lpi_j} = \width{L^{(\pitilde)}_j} , 
\end{equation*}
where the first and last equalities follow by construction. 
Our argument here centers on the \textit{reward} characterization of the width, where $\reward{S} = \sum_{e \in S}{- \log{ \pr{X_e > t}}}$. As explained in the definition of $\width{S}$, it is monotonically increasing in the reward of $S$. Thus, it suffices to show that
\begin{equation*}
 \reward{G - \Tstar_j} \geq \reward{\Lpi_j}.
\end{equation*}
We prove this in three short steps. First, since the reward is a modular set function, we may eliminate the common elements between $\Lpi_j$ and $G-\Tstar_j$, giving us $L'$ and $G'$ respectively. Second, we argue that, since $\Lpi_j$ is disjoint from $\Tstar_j$,
\begin{equation*}
    \Lpi_j \cap (G - \Tstar_j) = \Lpi_j \cap G,
\end{equation*}
and thus $L'$ must also be disjoint from $G$. Because $\greedyalg$ selects elements in order of decreasing reward density (i.e., in order of decreasing \textit{reward per unit cost}), it follows that every element in $G'$ has a higher reward density than every element in $L'$. From here, it suffices to show that $\cost{G'} \geq \cost{L'}$, or, since cost too is a modular set function, that 
\begin{equation*}
\cost{G - \Tstar_j} \geq \cost{\Lpi_j}.
\end{equation*} 
This third and final step follows from some basic facts about $G$, $\Tstar_j$, $\delta$, and $\Lpi_j$. First, we note that 
\begin{equation*}
    \cost{\Tstar_j} \leq B - \cost{\Lpi} + (\ell-1)\delta + \delta, 
\end{equation*}
where the first 3 terms follow from the definition of $\greedyalg$ and the final $\delta$ term follows from the fact that $\delta$ is the that maximum cost of any element.  From there, we note that 
\begin{equation*}
    \cost{G - \Tstar_j} = \cost{G} - \cost{\Tstar_j},
\end{equation*} 
since $\Tstar_j \subseteq G$. Once we note that $\cost{G} \geq B + \targetnew \delta$ by $\greedyalg$'s stopping condition, our proof of Lemma~\ref{lem:rect_covering} is complete, since
\begin{align*}
    \cost{G - \Tstar_j} &= \cost{G} - \cost{\Tstar_j}\\
    &\geq B + \targetnew \delta  - \cost{\Tstar_j}\\
    &\geq B + \targetnew \delta - \left( B - \cost{\Lpi} + \targetnew \delta \right)\\
    &= \cost{\Lpi}.
\end{align*}

\section{\texorpdfstring{The $\mink$-$\matroid$ Problem}{The MinK-Matroid Problem}}\label{sec:mink-mat}

We now turn our attention to a matroid-based variant of the $\mink$ problem, called the $\mink$-$\matroid$ problem. Here, the objective function $f(S)$ is again the sum of the smallest $k$ weights in $S$, but now the constraint is that the selection set $S$ must be an independent set in a given outer matroid $\Mouter = (\universe, \I)$, i.e., $\F = \I$. Our goal is to prove Theorem~\ref{thm:mink-mat}, which we restate now.
\begingroup
\def\thetheorem{\ref{thm:mink-mat}}
\begin{theorem}
    For the $\mink$-$\matroid$ problem where the random variables take values in $\intdomain$, there exists an adaptive policy that obtains a $\eightapprox$-approximation.
\end{theorem}
\addtocounter{theorem}{-1}
\endgroup
Recall the definition of $y_i(S)$ as ``the $i$-th smallest weight in $S$''. 
Following the derivation in \S~\ref{sec:mink-knap}, we can utilize the fact that $f(S) = \sum_{i=1}^k{y_i(S)}$ and apply Corollary~\ref{coro:sum-of-k} to reduce Theorem~\ref{thm:mink-mat} to finding an optimal policy for the $y_i$-$\F$ threshold problem. Fix the threshold $t$ and the index $i$. 
Recalling further the definition of $\trank{S} = \abs{\left \{ e \in S : X_e \leq t \right\}}$ and the equivalence of the two statements $y_i(S) \leq t$ and $\trank{S} \geq i$, we recover the $i$-th rank problem first treated in \S~\ref{sec:mink-knap},
\begin{equation*}
    \min_{\pi \in \adm{\F}}{\pr{\trank{\spi} \geq i}}.
\end{equation*}
We show that the standard matroid greedy algorithm $\matgreedy$ is optimal for this problem, and in doing so prove Theorem~\ref{thm:mink-mat}. A full description of $\matgreedy$'s use in our problem can be found in Algorithm~\ref{alg:mink-mat}.

\begin{algorithm}
\caption{$\matgreedy(\universe, \I, t)$}
\label{alg:mink-mat}
\begin{algorithmic}[1]
\State Initialization: $ S \gets \emptyset$; $\pool \gets \universe$

\While{$\left \{ e \in \universe - S \colon \brk{e} \cup S \in \I   \right \} \neq \emptyset$}
    \State $\ell \gets \argmax_{e \in \pool}{\pr{X_e \leq t}}$ \Comment{\textbf{Perform greedy selection}}
    \State $S \gets S \cup \brk{\ell}$
    \State $\pool \gets \pool - \brk{\ell}$
\EndWhile
\State \Return $S$
\end{algorithmic}
\end{algorithm}

\begin{lemma}\label{lem:mink-mat-opt}
    Given a matroid $\Mouter = (\universe, \I)$, let the set $G(\Mouter) = \matgreedy(\universe, \I, t)$. Then $G(\Mouter)$ is optimal for any rank problem with constraint set $\F = \I$, i.e., for any $i$,
    \begin{equation*}
        \pr{\trank{G(\Mouter)} \geq i} \geq \max_{\pi \in \adm{\F}}{\pr{\trank{\spi} \geq i}}.
    \end{equation*}
\end{lemma}

\subsection{Proof of Lemma~\ref{lem:mink-mat-opt}}
We prove this via induction on the matroid rank of the outer matroid $\Mouter$, i.e., the number of elements in a basis for $\Mouter$.
In the base case, the matroid rank of $\Mouter$ is $1$, and so only one element may be probed. Since only one element is probed without any additional information, all policies here are non-adaptive, and thus the optimal policy must be as well. Taking the element with the highest probability of being below-threshold is clearly optimal; this is precisely $\matgreedy$'s selection.

\begin{definition}
Define the contraction of  $\M$ by $S$ as
\begin{equation*}
    \contract{\M}{S} = \left \{T \subseteq \universe - S : T \cup S \in \I  \right \}.
\end{equation*}
In other words, the collection $\contract{\M}{S}$ is the collection of all sets which can be added to $S$ while maintaining independence. 
\end{definition}

We now treat the inductive case. 
Assume that, whenever the matroid rank of a given matroid $\M' = (\universe', \I')$ is smaller than the matroid rank of $\Mouter$, then the set $G(\M') = \matgreedy \left(\universe', \I', t\right)$ is an optimal policy for any rank problem with constraint $\I'$, i.e., that, for any $i$,
\begin{equation}\label{eq:mgreedy-inductive-assump}
            \pr{\trank{G(\M')} \geq i} \geq \max_{\pi \in \adm{\I'}}{\pr{\trank{\spi} \geq i}}.
\end{equation}
We show that $G(\Mouter)$ is optimal for any rank problem with constraint $\I$, i.e., that, for any $i$,
\begin{equation}\label{eq:mgreedy-opt}
\pr{\trank{G(\Mouter)} \geq i} \geq \max_{\pi \in \adm{\I}}{\pr{\trank{\spi} \geq i}}.
\end{equation}

Consider any $i$ and let $\pi^*_i$ be an optimal policy for the maximization problem on the right-hand-side of \eqref{eq:mgreedy-opt}.
Assume without loss of generality that the element $e$ is the first element probed by $\pi^*_i$ and let $\I'$ be the independent sets of $\contract{\Mouter}{\{e\}}$.
Then after seeing the realization of $X_e$, the rest of $\pi^*_i$ probes sets from $\I'$.
We use $\pi^*_i(x)$ to denote the rest of the policy $\pi^*_i$ after seeing $X_e=x$ for some $x\in\intdomain$.
Then $\pi^*_i(x)$ itself is a policy in $\adm{\I'}$.
Note that $\contract{\Mouter}{\{e\}}$ has strictly smaller matroid rank than $\Mouter$.
Therefore,
\begin{align}
\succprob{\ssup{\pi^*_i}}{i} &= \sum_{x \in \intdomain} {\pr{X_e = x}\pr{\trank{\ssup{\pi^*_i}} \geq i \given X_e = x}} \nonumber\\
&= \sum_{x \in \intdomain} {\pr{X_e = x}\succprob{\ssup{\pi^*_i(x)}}{i - \indc{x \leq t}}}\label{eq:mink-mat-secondline}\\
&\le \sum_{x \in \intdomain} {\pr{X_e = x}\succprob{G(\contract{\Mouter}{\brk{e}})}{i - \indc{x \leq t}}}\label{eq:mink-mat-indhyp}\\
&= \succprob{\brk{e}\cup G(\contract{\Mouter}{\brk{e}})}{i},\label{eq:mink-mat-opt}
\end{align}
where \eqref{eq:mink-mat-secondline} follows from the independence of the elements and \eqref{eq:mink-mat-indhyp} follows from the inductive hypothesis. Thus, we have shown that the non-adaptive set $\brk{e}\cup G(\contract{\Mouter}{\brk{e}})$ is optimal.


Now we specifically show that $G(\Mouter)=\matgreedy(\universe, \I, t)$ is optimal. Without loss of generality, let element $1$ be the element in $\universe$ with the largest probability of being below-threshold.  Then by the design of the algorithm $\matgreedy(\universe, \I, t)$, we have $G(\Mouter)=\{1\}\cup G(\contract{\Mouter}{\brk{1}})$. Note that
\begin{align}
&\mspace{17mu}\succprob{G(\Mouter)}{i}\nonumber\\
&=\succprob{\{1\}\cup G(\contract{\Mouter}{\brk{1}})}{i}\nonumber\\
&=\succprob{G(\contract{\Mouter}{\brk{1}})}{i} + \pr{\trank{G(\contract{\Mouter}{\brk{1}})} = i-1} \cdot \pr{X_1 \leq t}.\label{eq:rankGMO}
\end{align}
We next construct an independent set $S$ that contains element $1$ and elements from $\brk{e}\cup G(\contract{\Mouter}{\brk{e}})$, which will be used to lower bound \eqref{eq:rankGMO} with $\succprob{\brk{e}\cup G(\contract{\Mouter}{\brk{e}})}{i}$.
We start with $S=\{1\}$.  Then as long as $|S|<\left|\brk{e}\cup G(\contract{\Mouter}{\brk{e}})\right|$, by the matroid exchange property, there exists an element $e'$ in $\brk{e}\cup G(\contract{\Mouter}{\brk{e}})\setminus S$ such that $S\cup \{e'\}$ is still in $\I$.  We augment $S$ with this $e'$.  We continue this procedure until $|S|=\left|\brk{e}\cup G(\contract{\Mouter}{\brk{e}})\right|$.
By this construction, only one element in $\brk{e}\cup G(\contract{\Mouter}{\brk{e}})$ is not contained in $S$.  We denote this element as element $y$.
Clearly, $S\setminus\{1\}=\brk{e}\cup G(\contract{\Mouter}{\brk{e}})\setminus\{y\}$; let us call this set $L$, for readability.

We next lower bound \eqref{eq:rankGMO}. Note that $\pr{X_1 \leq t}\ge \pr{X_y\le t}$ by definition.  Now let us focus on $\trank{G(\contract{\Mouter}{\brk{1}})}$.  Since the rank of the matroid $\contract{\Mouter}{\brk{1}}$ is smaller than that of $\Mouter$, then by the inductive assumption in \eqref{eq:mgreedy-inductive-assump}, we have
\begin{gather*}
\succprob{G(\contract{\Mouter}{\brk{1}})}{i}\ge \succprob{L}{i}.
\end{gather*}
Taking $x = \succprob{G(\contract{\Mouter}{\brk{1}})}{i}$ and $y = \succprob{L}{i}$, we note that
\begin{equation*}
    x + (1-x)\cdot \pr{X_1 \leq t} \geq y + (1-y)\cdot \pr{X_1 \leq t}.
\end{equation*}

Therefore, the terms in \eqref{eq:rankGMO} can be lower bounded as
\begin{align*}
&\mspace{17mu}\succprob{G(\Mouter)}{i}\\
&=\succprob{\brk{1} \cup G(\contract{\Mouter}{\brk{1}}}{i}\\
&=\succprob{G(\contract{\Mouter}{\brk{1}})}{i} + \pr{\trank{G(\contract{\Mouter}{\brk{1}})} = i-1} \cdot \pr{X_1 \leq t}\\
&\geq\succprob{L}{i} + \pr{\trank{L} = i-1} \cdot \pr{X_1 \leq t}\\
&\ge \succprob{L}{i} + \pr{\trank{L} = i-1} \cdot \pr{X_y \leq t}\\
&=\succprob{\brk{y}\cup L}{i}\\
&=\succprob{\brk{e}\cup G(\contract{\Mouter}{\brk{e}})}{i}.
\end{align*}
Since we have already shown that $\brk{e}\cup G(\contract{\Mouter}{\brk{e}})$ is optimal in \eqref{eq:mink-mat-opt}, this completes the proof that $G(\Mouter)=\matgreedy(\universe, \I, t)$ is optimal.

\section{\texorpdfstring{$\minbasis$-$\card$ Problem}{The MinBasis-Cardinality Problem}}\label{sec:minbasis-card}

In this section, we focus on the $\mink$-$\cardinality$ problem, where, given an inner matroid $\Minner = (\universe, \I)$ and budget $B$, the objective function $f(S)$ is the weight of the smallest basis for $\Minner$ contained in $S$ and the constraint is a cardinality constraint, i.e., $\F = \left \{ S \subseteq \universe \colon \abs{S} \le B \right \}$. In our notation, the $\mink$-$\cardinality$ problem can be written as:
\begin{equation}\label{eq:opt-mink-cardinality}
\min_{\pi \in \adm{\F}}{\ex{\min_{\substack{T \subseteq \spi\\ \text{$T$ is a basis for $\Minner$}}}{\sum_{e \in T}{X_e}}}}.
\end{equation}
Our goal is proving Theorem~\ref{thm:minbasis-card}, which we restate now.
\begingroup
\def\thetheorem{\ref{thm:minbasis-card}}
\begin{theorem}
    For the $\minbasis$-$\cardinality$ problem where the random variables takes values in $\intdomain$, there exists an adaptive policy that obtains a $\minbasiscardapprox$-approximation.
\end{theorem}
\addtocounter{theorem}{-1}
\endgroup

The proof of Theorem~\ref{thm:minbasis-card} is again through Corollary~\ref{coro:sum-of-k}.
However, it is not obvious how to write $f(S)$ as $\sum_{i=1}^k{g_i(S)}$ for a monotonic function sequence $(g_i)$ where each $g_i$ is a non-decreasing function taking values in $\intdomain$.
In our proof, we first show that we can specify $g_i(S)$ to be the $i$-th smallest element in the minimum-weight basis contained in $S$ generated by the matroid greedy algorithm.
Then we show inductively that $\adapgreedy$, a simple adaptive matroid greedy algorithm described in Algorithm~\ref{alg:minbasis-card}, is optimal for a rank problem that is equivalent to the resulting $g_i$-$\cardinality$ threshold problem.

\subsection{Reduction to rank problems}
\paragraph{Decomposition of the objective function.}
We first describe the decomposition $f(S) = \sum_{i=1}^k{g_i(S)}$. Given a set $S$ and a weighted matroid $\Minner$, it is well known that the matroid greedy algorithm will find a basis for $\Minner$ of minimum weight, where the matroid greedy algorithm at each step adds a minimum-weight element among the elements whose addition would preserve the independence. Let $G(S, \Minner)$ be the set generated by this process. It follows that for the selection set $\spi$ of any policy $\pi$,
\begin{equation}\label{eq:minbasis-reinterp}
    \min_{\substack{T \subseteq \spi\\ \text{$T$ is a basis for $\Minner$}}}{\sum_{e \in T}{X_e}} = \sum_{e \in G\left(\spi, \Minner\right)}{X_e}.
\end{equation}
In other words, to minimize our original objective, it suffices to minimize the cumulative weight of the post-selection greedy set $G\left(\spi, \Minner\right)$. Note also that, by ordering the summation of elements within this greedy set accordingly, we have
\begin{equation*}
    \sum_{e \in G\left(\spi, \Minner\right)}{X_e} = \sum_{i=1}^k{y_i\left(G\left(\spi, \Minner\right) \right)},
\end{equation*}
where one must recall that $y_i(T)$ is the weight of the $i$-th smallest element in the set $T$.
Then we specify the function $g_i$ in the decomposition $f(S) = \sum_{i=1}^k{g_i(S)}$ as $g_i(S) = y_i(G(S, \Minner))$. To complete our reduction, we show that these $g_i$'s satisfy the conditions of Corollary~\ref{coro:sum-of-k}, i.e., the function sequence $(g_i)$ is monotonic, and each function $g_i$ is a non-decreasing function taking values in $\intdomain$. 

First, we show that $g_i(S) \leq g_{i+1}(S)$. By the nature of greedy selection, elements are chosen for $G(S,\Minner)$ in order of increasing weight. It follows that the $i$-th smallest element in $G(S, \Minner)$ must be smaller than the $(i+1)$-th smallest element in $G(S, \Minner)$.

Second, we show that for each function $g_i$, given a subset $T \subseteq S$, $g_i(S) \leq g_i(T)$. It suffices to show that for all $t\in\intdomain$, if $g_i(T) \leq t$, then $g_i(S) \leq t$. To see this, note that if $g_i(T) \leq t$, then there must exist an independent set of size $i$ within $T$ within which all elements are of weight at most $t$. We again argue via the greedy selection process. Note that $S$ must also contain the size $i$ independent set previously mentioned. It follows from the matroid exchange axiom that, in the $\ell$-th round of greedy selection for $\ell \leq i$, there must always exist an element of weight $\leq t$ that we can add to the greedy set we are building. As such, the the first $i$ elements in our greedy set must have weight $\leq t$, and thus $g_i(S) \leq t$.

Third, since every weight $X_e$ takes values $\intdomain$, it follows that $g_i(S)$ takes values in $\intdomain$. Thus, the conditions of Corollary~\ref{coro:sum-of-k} are satisfied, and to prove Theorem~\ref{thm:minbasis-card}, it suffices to exhibit an optimal algorithm for the $g_i$-$\F$ threshold problem. Before that, we introduce a helpful reinterpretation of this problem by defining a rank problem.

\paragraph{From threshold problems to rank problems.}
\begin{definition}[The matroid threshold rank $\Mtrank{\Minner}{S}$]
Given a threshold $t$ and a matroid $\Minner$, for any set $S$, let $\Mtrank{\Minner}{S}$ denote the size of the largest independent set contained in $S$ within which all elements are below the threshold $t$.
Using our previous definition $\trank{T}=\left|\{e\in T\colon X_e\le t\}\right|$, we can define $\Mtrank{\Minner}{S}$ as
\begin{equation*}
    \Mtrank{\Minner}{S} = \max_{T \subseteq S, T \in \Minner}{\trank{T}}.
\end{equation*}
We call this the matroid \emph{threshold} rank to distinguish it from the matroid rank of $\Minner$, which is simply the size of a basis for $\Minner$.
\end{definition}
Following the exposition of $\mink$-$\knapsack$ problem, we also have the following equivalence between the threshold condition and a rank condition: $g_i(S) \leq t$ if and only if $\Mtrank{\Minner}{S} \geq i$. To see this, if $g_i(S) \leq t$, then $S$ must contain an independent set of size $i$ wherein all elements are of weight $\leq t$. For the other direction, if $\Mtrank{\Minner}{S} \geq i$, then there exists an independent set $T \subseteq S$ with $\trank{T}\ge i$.  Since $T$ is an independent set itself, we have $G\left(T, \Minner\right)=T$ and thus $g_i(T)\le t$.  Therefore, $g_i(S) \leq g_i(T)\le t$ by the property that $g_i(\cdot)$ is a non-increasing function.

By this equivalence, it follows that the $g_i$-$\F$ threshold problem satisfies
\begin{equation}\label{eq:minbasis-card-thresh}
    \min_{\pi \in \adm{\F}}{\pr{g_i(S) \leq t}} = \min_{\pi \in \adm{\F}}{\pr{\Mtrank{\Minner}{\spi}\ge i}}.
\end{equation}
We use the above characterization in the remainder of this section, which we devote to proving the following lemma.
\begin{lemma}\label{lem:minbasis-opt}
    The adaptive greedy algorithm $\adapgreedy$, described in Algorithm~\ref{alg:minbasis-card}, is optimal for the rank problem in \eqref{eq:minbasis-card-thresh} for all $i$.
\end{lemma}

\newcommand{\useful}{\textsc{Useful}}

\begin{algorithm}
\caption{$\adapgreedy(\universe, \I, B, t)$}
\label{alg:minbasis-card}
\begin{algorithmic}[1]
\State Initialization: $ S \gets \emptyset$; $\useful \gets \left \{e \in \universe \colon \brk{e} \in \I  \right \}$; $T \gets \emptyset$

\While{$\useful \neq \emptyset$ and $\abs{S} < B$}
    \State $\ell \gets \argmax_{e \in \useful}{\pr{X_e \leq t}}$ \Comment{\textbf{Perform greedy selection}}
    \State $S \gets S \cup \brk{\ell}$
    \If{ $X_\ell \leq t$ }
        \State $T \gets T \cup \brk{\ell}$
    \EndIf
    \State $\useful \gets \left \{e \in \universe - S \colon  \brk{e} \cup T \in \I  \right \}$ \Comment{\textbf{Adaptively update useful set}}
\EndWhile
\State \Return $S$
\end{algorithmic}
\end{algorithm}

\subsection{Proof of Lemma~\ref{lem:minbasis-opt}}
We prove Lemma~\ref{lem:minbasis-opt} via induction on the budget $B$. 

In the base case, let the budget $B \leq i$. First, note that, if $B < i$, then no policy can succeed. When the required number of below-threshold elements ($i$) is equal to the number of elements one can probe ($B$), then, in order to reach that threshold $i$, every probe must end up below-threshold. Moreover, in order to succeed every element $e$ selected must be \emph{useful}: when added to the selection set $S$, the new set $S\cup \brk{e}$ must be independent. Now, let the element $e$ be the first element probed by a policy $\pi$, and $\pi_x$ be the policy that describes $\pi$'s behavior in the case when $X_e = x$. Then we have
\begin{align*}
    \pr{\Mtrank{\M}{\spi} \geq i} &= \sum_{x \in \intdomain}{\pr{X_e = x}\pr{\Mtrank{\M}{\spix\cup \brk{e}} \geq i \given X_e = x}}\\
    &= \sum_{x \leq t}{\pr{X_e = x}\pr{\Mtrank{\M}{\spix\cup \brk{e}} \geq i \given X_e = x}}.
\end{align*}
To proceed, we make use of the following claim.
\newcommand{\Isup}[1]{\I_{brk{#1}}}
\begin{claim}\label{clm:minb-recurse}
     If $X_e \leq t$, then $\Mtrank{\Minner}{\spix\cup\brk{e}} \geq i$ if and only if $\Mtrank{\contract{\Minner}{\brk{e}}}{\spix} \geq i - 1$.
\end{claim}
\begin{proof}
    Given a matroid $\M'= \left(\universe', \I' \right)$, say that a set $S \in \M'$  if $S \in \left\{T\subseteq \universe' : T \in \I' \right \}$. Specifically, for the element $e$ and the matroid contraction $\contract{\Minner}{\brk{e}}$, a set $S$ lies in $\contract{\Minner}{\brk{e}}$ if and only if $S$ does not contain the element $e$ and $S \cup \brk{e} \in \Minner$.

    We first show that $\Mtrank{\Minner}{\spix \cup \brk{e}} \geq i$ implies that $\Mtrank{\contract{\Minner}{\brk{e}}}{\spix} \geq i - 1$. Call a set $S$ below-threshold if every element $v$ in $S$ has $X_v \leq t$, and recall that
     \begin{equation*}
        \Mtrank{\contract{\Minner}{\brk{e}}}{\spix}= \max_{S \subseteq \spix, S \in \contract{\Minner}{\brk{e}}}{\trank{S}}.
    \end{equation*}
    It follows that, if $\Mtrank{\Minner}{\spix \cup \brk{e}} \geq i$, then, by definition, the set $\spix \cup \brk{e}$ contains a below-threshold independent set of size $j$; call this set $T$. We now show that $\spix$ contains a below-threshold set $T' \in \contract{\Minner}{\brk{e}}$ of size $i-1$. Without loss of generality, assume $i > 1$. Since 
    \begin{equation*}
        \abs{T} = i  > 1 = \abs{\brk{e}},
    \end{equation*} 
    the matroid exchange axiom guarantees that there must exist an element $v_1 \in T$ such that the combined set $\brk{v_1, e} \in \Minner$.  Continuing this process from $x = 2$ to $x=i-1$, we find elements $v_x \in T$ such that the combined set $\left (\brk{e} \cup \bigcup_{r=1}^{x-1}{v_r} \right ) \cup v_x \in \Minner$.  Now, observe that the augmenting set $T' = \bigcup_{r=1}^{i-1}{v_r}$ is in $\contract{\Minner}{\brk{e}}$, and that, since $T'$ is a subset of the below-threshold set $T$, the augmenting set $T'$ itself must also be below-threshold. Thus,
    \begin{equation*}
        \Mtrank{\contract{\Minner}{\brk{e}}}{\spix} \geq \trank{T'} = i - 1.
    \end{equation*}
    
    The other direction is easy. If $\Mtrank{\contract{\Minner}{\brk{e}}}{\spix} \geq i - 1$, then there must exist a size $(i-1)$ below-threshold set $T'$ contained in $\spix$, which lies in $\contract{\M}{\brk{e}}$. It follows that $T'\cup \brk{e} \in \Minner$, and, since $X_e \leq t$, the combined set $T'\cup \brk{e}$ is also below-threshold. Thus, we have 
    \begin{equation*}
        \Mtrank{\Minner}{\spix \cup \brk{e}} \geq \trank{T' \cup \brk{e}} = i,
    \end{equation*}
    as desired.
    
\end{proof}

Continuing from the claim and recalling that $\pi_x$ represents the continuing behavior of the policy $\pi$ after it probes the element $e$ and obtains $X_e = x$, we see that
\begin{align*}
    \pr{\Mtrank{\M}{\spi} \geq i}  &= \sum_{x \leq t}{\pr{X_e = x}\pr{\Mtrank{\M}{\spix\cup \brk{e}} \geq i \middle | X_e = x}}\\
    &= \sum_{x \leq t}{\pr{X_e = x}\pr{\Mtrank{\contract{\Minner}{\brk{e}}}{\spix} \geq i-1}}\\
    &\leq \sum_{x \leq t}{\pr{X_e = x}\max_{\pi' \in \adm{\F_{\brk{e}}}}{\pr{\Mtrank{\contract{\Minner}{\brk{e}}}{\ssup{\pi'}} \geq i-1}}}\\
    &= \pr{X_e \leq t} \max_{\pi' \in \adm{\F_{\brk{e}}}}{\pr{\Mtrank{\contract{\Minner}{\brk{e}}}{\ssup{\pi'}} \geq i-1}},
\end{align*}
where $\F_{\brk{e}} = \left \{ S \subseteq \universe - \brk{e}\colon   \abs{S} \leq B - 1 \right \}$. Continuing the same procedure (in downwards induction), we find that, when $B=i$, any optimal adaptive policy has the same success probability as a non-adaptive policy, i.e., that
\begin{align*}
    \max_{\pi \in \adm{\F}}{\pr{\Mtrank{\Minner}{\spi} \geq i}} &= \max_{S \in \F}{\pr{\Mtrank{\Minner}{S} \geq i}}\\
    &= \max_{S \in \F, S \in \M}{\prod_{\ell \in S}{\pr{X_\ell \leq t}}}\intertext{which after some manipulations becomes}
    &= \exp{ \left(\max_{S \in \F, S \in \M}{\sum_{\ell \in S}{\log{\left[\pr{X_\ell \leq t}\right] }}} \right )}.
\end{align*}
But $\max_{S \in \F, S \in \M}{\sum_{\ell \in S}{\log{\left[\pr{X_\ell \leq t}\right] }}}$ is a max-weight basis problem for the matroid $(\universe, \left \{ S \in \I : \abs{S} \leq B \right \})$ with element weights $\log{\pr{X_e \leq t}}$. It follows that the matroid greedy algorithm is optimal for this problem, and thus that the adaptive greedy algorithm is optimal (since, if every element probed is below-threshold, then the adaptive greedy algorithm is equivalent to the matroid greedy algorithm). This completes the base case.

\paragraph{Inductive Case.} Now, assume inductively that, for all inner matroids $\Minner' = (\universe', \I')$ and budgets $B' < B$, the adaptive greedy algorithm $\adapgreedy(\universe', \I', B' , t)$ is optimal for all targets $i'$. We split the proof of the inductive case into two claims. First, we show that, under our inductive assumption, it is always optimal to follow the adaptive greedy policy after probing the one's first element. Second, we show that the element first probed by the adaptive greedy strategy is optimal. It follows directly from these two claims that the adaptive greedy strategy is optimal. Thus, we now state and prove the claims.
\begin{claim}\label{clm:minb-after}
Assume that, for all inner matroids $\Minner' = (\universe', \I')$, budgets $B' < B$, and targets $i'$, the adaptive greedy algorithm $\adapgreedy(\universe', \I', B', t)$ is optimal for the problem \begin{equation*}
    \min_{\pi \in \adm{\F'}}{\pr{\Mtrank{\Minner'}{\spi}} \geq i'}.
\end{equation*} Let $\F = \left \{ S \subseteq \universe \middle | \abs{S} \leq B \right \}$. For any policy $\pi \in \adm{\F}$, let $e$ be the first element $\pi$ probes. If $\pi'$ is the policy that probes $e$ then follows the adaptive greedy strategy, then
\begin{equation*}
     \pr{\Mtrank{\M}{\spi} \geq i} \leq  \pr{\Mtrank{\M}{\ssup{\pi'}} \geq i}.
\end{equation*}
\end{claim}
\begin{proof}
Let $\AG(-) = \adapgreedy(\universe - \brk{e}, \I, B-1, t)$ and $\AG(+) = \adapgreedy(\universe - \brk{e}, \I_{\brk{e}}, B-1, t)$, be the adaptive greedy procedure which would be optimal follow after probing element $e$, given that either $X_e > t$ or $X_e \leq t$, respectively. Then, using Claim~\ref{clm:minb-recurse},
\begin{align*}
    \pr{\Mtrank{\M}{\spi} \geq i}  &= \sum_{x \in \intdomain}{\pr{X_e = x}\pr{\Mtrank{\M}{\spix\cup \brk{e}} \geq i \middle | X_e = x}}\\
    &= \sum_{x \le t}{\pr{X_e = x}\pr{\Mtrank{\M}{\spix\cup \brk{e}} \geq i \middle | X_e = x}}\\
    &\;\;+ \sum_{x > t}{\pr{X_e = x}\pr{\Mtrank{\M}{\spix\cup \brk{e}} \geq i \middle | X_e = x}}\\
    &= \sum_{x \le t}{\pr{X_e = x}\pr{\Mtrank{\contract{\Minner}{\brk{e}}}{\spix} \geq i -1}}\\ 
    &\;\;+ \sum_{x > t}{\pr{X_e = x}\pr{\Mtrank{\M}{\spix} \geq i}}\\
    &\leq \sum_{x \le t}{\pr{X_e = x}\pr{\Mtrank{\contract{\Minner}{\brk{e}}}{\ssup{\AG(-)}} \geq i -1}}\\
    &\;\;+ \sum_{x > t}{\pr{X_e = x}\pr{\Mtrank{\M}{\ssup{\AG(+)}} \geq i}}\\
    &= \pr{X_e \leq t}\pr{\Mtrank{\contract{\Minner}{\brk{e}}}{\ssup{\AG(+)}} \geq i -1} \\
    &\;\;+ \pr{X_e > t}\pr{\Mtrank{\M}{\ssup{\AG(-)}} \geq i},
\end{align*}
where the last line is precisely the probability of success if one probes $e$ then follows the adaptive greedy strategy.
\end{proof}

We now show that the adaptive greedy algorithm does not choose a sub-optimal first element.
\begin{claim}\label{clm:minb-first}
Let element $1$ be the element that would be picked first by the adaptive greedy policy. Then, for the threshold problem \eqref{eq:minbasis-card-thresh}, there exists an optimal policy which probes $1$ first.
\end{claim}
\begin{proof}
    Assume there exists a policy $\pi$ whose first selection is an element $e$ which is not $1$. We will show how to construct a policy $\pi'$ which first probes $1$, and succeeds at least as often as $\pi$. There are two cases, depending on whether or not $\brk{e,1} \in \I$.

 In the first case, $\brk{e, 1} \in \I$. From the above, we know that it is optimal to 
follow $\AG$ after probing $e$. Regardless of the value of $X_e$, the next element to be probed would then be $1$, since 
\begin{equation*}
    1 = \argmax_{\ell \in \universe}{\pr{X_\ell \leq t}} = \argmax_{\ell \in \universe - \brk{e}, \brk{e, \ell} \in \I}{\pr{X_\ell \leq t}}.
\end{equation*}
From Claim~\ref{clm:minb-after}, we know that, after probing element $1$ and element $e$, it is optimal to follow the adaptive greedy strategy. Since we always probe both element $e$ and element $1$ and follow the same strategy, we can switch the order that we probe them in without changing the probability of success. In other words, there must exist an optimal policy which probes element $1$ first (then element $e$, then follows $\AG$), as desired.

In the second case, $\brk{e, 1} \not \in \I$. Assume (without loss of success probability, by Claim~\ref{clm:minb-recurse}) that $\pi$ follows the adaptive greedy algorithm. We note three facts. First, note that, if $\brk{e, 1} \not \in \I$, then $\contract{\Minner}{\brk{e}} = \contract{\Minner}{\brk{1}}$, since if there exists $T$ such that $T \cup \brk{e} \in \I$, then, after augmenting $\brk{1}$ with $T \cup \brk{e}$, we must obtain $T \cup \brk{1}$ and vice versa (since independence is subset-closed). Second, note that, if $X_e > t$, then the second element probed would be element $1$ since $1 = \argmax_{\ell \in \universe - \brk{e}}{\pr{X_\ell \leq t}}$. For brevity, we write $\AG$ as an abbreviation of $\adapgreedy$. Third, note that we can always eliminate non-independent elements from the universe of the greedy argument, i.e., if $\brk{\ell} \not \in \I'$, then $\AG(\universe', \I', B', t) = \AG(\universe' - \brk{\ell}, \I', B', t)$. From this third fact along with the first, it follows that
\begin{equation*}
    \AG(\universe - \brk{e}, \I_{\brk{e}}, B', t) = \AG(\universe - \brk{e,1}, \I_{\brk{e}}, B', t) = \AG(\universe - \brk{1}, \I_{\brk{1}},B', t),
\end{equation*}
where one should recall that $\I_{\brk{v}}$ is the contraction of $\I$ by $\brk{v}$, i.e., $\I_{\brk{v}} = \left \{ S \in \I - \brk{v}: S \cup \brk{v} \in \I  \right \}$ 

We consider the following policy $\pi'$: First, probe element $1$. If $X_1 \leq t$, then do what $\pi$ does when $X_e \leq t$ (follow $\AG(\universe - \brk{e}, \I_{\brk{e}}, t)$). If $X_1 > t$, then probe element $e$ and, if $X_e \leq t$, do what $\pi$ does when $X_e > t$ and $X_1 \leq t$; otherwise, do what $\pi$ does when both $X_e > t$ and $X_1 > t$.

It suffices to show that $\pr{\Mtrank{\Minner}{\ssup{\pi'}} \geq i}- \pr{ \Mtrank{\Minner}{\spi} \geq i} \geq 0$. By construction, we see that
\begin{align*}
    \pr{\Mtrank{\Minner}{\spi} \geq i \middle | X_e \leq t} &= \pr{\Mtrank{\contract{\Minner}{\brk{e}}}{\ssup{\AG(\universe - \brk{e}, \I_{\brk{e}},B-1, t)}} \geq i-1}\\
    &= \pr{\Mtrank{\contract{\Minner}{\brk{1}}}{\ssup{\AG(\universe - \brk{1}, \I_{\brk{1}},B-1, t)}} \geq i-1}\\
    &= \pr{\Mtrank{\Minner}{\ssup{\pi'}} \geq i \middle | X_1 \leq t};
\end{align*}
we assign this value a shorthand $\Ra$.
Similarly,
\begin{align*}
    \pr{\Mtrank{\Minner}{\spi} \geq i \middle | X_e > t, X_1 \leq t} &= \pr{\Mtrank{\Minner}{\ssup{\AG(\universe - \brk{e,1}, \I_{\brk{e}}, B-2, t)}} \geq i  -1}\\
    &= \pr{\Mtrank{\Minner}{\ssup{\AG(\universe - \brk{e,1}, \I_{\brk{1}}, B-2, t)}} \geq i  -1}\\
    &= \pr{\Mtrank{\Minner}{\ssup{\pi'}} \geq i \middle | X_1 \leq t, X_e > t},
\end{align*}
which we call $\Rb$. And, likewise, we have
\begin{equation*}
    \pr{\Mtrank{\Minner}{\spi} \geq i \middle | X_e > t, X_1 > t} = \pr{\Mtrank{\Minner}{\ssup{\pi'}} \geq i \middle | X_e > t, X_1 > t},
\end{equation*}
which we call $\Rc$.
Putting these observations together, we find that, abbreviating $\pr{X_e \leq t}$ as $p_e$,
\begin{align*}
    \pr{ \Mtrank{\Minner}{\spi} \geq i} &= \pr{X_e \leq t} \Ra +
    \pr{X_e > t}\left(\pr{X_1 \leq t}\Rb + \pr{X_1 > t}\Rc \right)\\
    &= p_e \Ra + (1-p_e)\left(p_1 \Rb + (1-p_1) \Rc \right),
\end{align*}
and analogously that 
\begin{equation*}
    \pr{\Mtrank{\Minner}{\ssup{\pi'}} \geq i} = p_1 \Ra + (1-p_1) \left(p_e \Rb + (1- p_e)\Rc \right).
\end{equation*}

Comparing the two, we find
\begin{align*}
    \pr{\Mtrank{\Minner}{\ssup{\pi'}} \geq i}- \pr{ \Mtrank{\Minner}{\spi} \geq i} &= (p_1 - p_e) (\Ra - \Rb).
\end{align*}
Since $p_1 \geq p_e$ by definition, it suffices to show that $\Ra \geq \Rb$, i.e., that 
\begin{equation*}
    \pr{\Mtrank{\contract{\Minner}{\brk{1}}}{\ssup{\AG(\universe - \brk{1}, \I_{\brk{1}}, B-1, t)}} \geq i-1} \geq \pr{\Mtrank{\Minner}{\ssup{\AG(\universe - \brk{e,1}, \I_{\brk{1}}, B-2, t)}} \geq i  -1},
\end{equation*} but, from a simple coupling argument, one sees that having an additional element available for probing and having a larger budget to probe can only increase the probability of the adaptive greedy algorithm's success. In other words, $\adapgreedy \left(\universe - \brk{1}, \I_{\brk{1}}, B-1, t\right)$ will succeed more often than $\AG(\universe - \brk{e,1}, \I_{\brk{1}}, B-2, t)$. Thus, $\Ra \geq \Rb$, and
\begin{equation*}
    \pr{\Mtrank{\Minner}{\ssup{\pi'}} \geq i}- \pr{ \Mtrank{\Minner}{\spi} \geq i} \geq 0,
\end{equation*}
as desired. Since $\pi$ is optimal and $\pi' \in \adm{\F}$, it follows that $\pi'$ is optimal as well, meaning that, in the case where $\brk{e,1} \not \in \I$, there exists an optimal policy which probes element $1$ first. This completes the proof of Claim~\ref{clm:minb-first}, and thus our proof of Lemma~\ref{lem:minbasis-opt} and, accordingly, our proof of Theorem~\ref{thm:minbasis-card}.
\end{proof}


\appendix

\section{Adaptivity Gap Example}\label{sec:appendix}
\begin{example}\label{ex:adapgap}
    Consider the following probing problem. We are given a universe of three elements,  $X_1$, $X_2$, and $X_3$ to choose from, independently distributed from each other, with distributions we describe later.  We are allowed to choose 2 of these random variables for our set $S$, and the end objective $f(S)$ is the minimum weight element in $S$. The distributions are as follows:  $X_1$ is $N^2$ with probability (w.p.) $\frac{1}{N^2}$ and $1$ otherwise, $X_2$ is $N^2$ w.p.\ $\frac{1}{N}$ and $0$ otherwise, and $X_3$ is $N$ w.p.\ $1$.
    
    Computing the expected value of each fixed two-element set, we find that
\begin{align*}
    \ex{f(\{1,3\})} &= \ex{\min(X_1, N)} = \left(1 - \frac{1}{N^2} \right)\cdot 1 + \left (\frac{1}{N^2} \right)\cdot N \geq 1,\\
    \ex{f(\{1,2\})} &= \left(1 - \frac{1}{N} \right)\ex{X_1} = \left(1 - \frac{1}{N} \right)\left [ \left(1 - \frac{1}{N^2} \right)\cdot 1 + \left(\frac{1}{N^2} \right)\cdot N^2 \right ] \geq 1,\\
    \ex{f(\{2,3\})} &= \ex{\min(X_2, N)} = \frac{1}{N}\cdot N = 1.
\end{align*}

    Now consider the following adaptive policy, which we call $\pi$: First, probe element $1$. If $X_1 = 1$, then probe element $2$; otherwise probe element $3$. Intuitively, this policy probes the `risky’ element 2  only if it has already secured an objective value of $1$. Computing the expected objective value (taking $\spi$ to be the set chosen by policy $\pi$), we find
\begin{equation*}
    \ex{f\left(\spi\right)} = \left(1 - \frac{1}{N^2} \right)\cdot  \frac{1}{N}\cdot 1 + \left(\frac{1}{N^2} \right) \cdot N \leq \frac{2}{N}.
\end{equation*}
    Thus, the adaptivity gap of this instance is at least $\frac{N}{2}$, which can be made arbitrarily large.
  \end{example}


\bibliographystyle{abbrvnat}
\bibliography{refs,probe}

\end{document}